\documentclass[aps,prb,twocolumn,superscriptaddress,floatfix]{revtex4-1}

\usepackage{graphicx}% Include figure files
\usepackage{dcolumn}% Align table columns on decimal point
\usepackage{bm}% bold math
\usepackage{times}% font that prx use
\usepackage{color}
\begin{document}
\bibliographystyle{apsrev4-1}

\title{Extremely Large Magnetoresistance and Electronic Structure of TmSb}

\author{Yi-Yan Wang}
\affiliation{Department of Physics and Beijing Key Laboratory of Opto-electronic Functional Materials $\&$ Micro-nano Devices, Renmin University of China, Beijing 100872, P. R. China}

\author{Hongyun Zhang}
\affiliation{State Key Laboratory of Low Dimensional Quantum Physics and Department of Physics, Tsinghua University, Beijing 100084, P.R. China}

\author{Xiao-Qin Lu}
\affiliation{Department of Physics and Beijing Key Laboratory of Opto-electronic Functional Materials $\&$ Micro-nano Devices, Renmin University of China, Beijing 100872, P. R. China}

\author{Lin-Lin Sun}
\affiliation{Department of Physics and Beijing Key Laboratory of Opto-electronic Functional Materials $\&$ Micro-nano Devices, Renmin University of China, Beijing 100872, P. R. China}

\author{Sheng Xu}
\affiliation{Department of Physics and Beijing Key Laboratory of Opto-electronic Functional Materials $\&$ Micro-nano Devices, Renmin University of China, Beijing 100872, P. R. China}

\author{Zhong-Yi Lu}
\affiliation{Department of Physics and Beijing Key Laboratory of Opto-electronic Functional Materials $\&$ Micro-nano Devices, Renmin University of China, Beijing 100872, P. R. China}

\author{Kai Liu}
\affiliation{Department of Physics and Beijing Key Laboratory of Opto-electronic Functional Materials $\&$ Micro-nano Devices, Renmin University of China, Beijing 100872, P. R. China}

\author{Shuyun Zhou}
\affiliation{State Key Laboratory of Low Dimensional Quantum Physics and Department of Physics, Tsinghua University, Beijing 100084, P.R. China}
\affiliation{Collaborative Innovation Center of Quantum Matter, Beijing 100084, P.R. China}

\author{Tian-Long Xia}\email{tlxia@ruc.edu.cn}
\affiliation{Department of Physics and Beijing Key Laboratory of Opto-electronic Functional Materials $\&$ Micro-nano Devices, Renmin University of China, Beijing 100872, P. R. China}

\date{\today}
\begin{abstract}
We report the magneto-transport properties and the electronic
structure of TmSb. TmSb exhibits extremely large transverse
magnetoresistance and Shubnikov-de Haas (SdH) oscillation at
low temperature and high magnetic field. Interestingly, the split
of Fermi surfaces induced by the nonsymmetric spin-orbit interaction
has been observed from SdH oscillation. The analysis of the
angle-dependent SdH oscillation illustrates the contribution of each
Fermi surface to the conductivity. The electronic structure revealed by angle-resolved photoemission spectroscopy (ARPES) and
first-principles calculations demonstrates a gap at $X$ point and the
absence of band inversion. Combined with the trivial Berry phase
extracted from SdH oscillation and the nearly equal concentrations
of electron and hole from Hall measurements, it is suggested that
TmSb is a topologically trivial semimetal and the observed XMR
originates from the electron-hole compensation and high mobility.
\end{abstract} \maketitle

\setlength{\parindent}{1em}

\section{Introduction}

Recently, rare earth monopnictides LnX (Ln=La, Y, Ce, Nd and X=Sb,
Bi) have drawn much attention and been studied
widely\cite{zeng2015topological,tafti2015resistivity,tafti2016temperature,PhysRevLett.117.127204,Ban2017observation,
PhysRevB.96.081112,PhysRevB.96.125112,sun2016large,PhysRevB.93.241106,PhysRevB.94.081108,PhysRevB.95.115140,nayak2017multiple,
Singha2017Fermi,PhysRevB.93.235142,PhysRevB.94.165163,ghimire2016magnetotransport,Yu2017Magnetoresistance,pavlosiuk2016giant,
PhysRevLett.117.267201,PhysRevB.96.075159,alidoust2016new,ye2017extreme,PhysRevB.96.041120,kuroda2017experimental,PhysRevB.93.205152,neupane2016observation,PhysRevB.96.035134}.
In these materials, extremely large magnetoresistance (XMR) is a
remarkable signature since conventional nonmagnetic metals usually
show a small magnetoresistance (MR) of only a few percent. XMR has
also been observed in several other materials such as
WTe$_2$\cite{ali2014large,PhysRevB.92.180402} and
(Nb/Ta)As$_2$\cite{PhysRevB.94.041103,PhysRevB.93.184405,wu2016giant,luo2016anomalous,PhysRevB.93.195119}.
Several mechanisms have been proposed to explain the origin of XMR,
for example, magnetic field induced metal-to-insulator
transition\cite{tafti2015resistivity}, the breaking of topological
protection\cite{liang2015ultrahigh} or the compensation of hole and
electron\cite{sun2016large,PhysRevB.93.235142}. For a semimetal with
topologically nontrivial electronic structure, the topological
protection suppresses backscattering at zero magnetic field. The
application of a field will break the protection and result in
XMR\cite{liang2015ultrahigh}. However, nontrivial topological state
is not indispensable for the generation of XMR since topologically
trivial materials (such as LaSb\cite{PhysRevLett.117.127204},
YSb\cite{PhysRevLett.117.267201,Yu2017Magnetoresistance}, and
CeSb\cite{PhysRevB.96.041120}) can also exhibit XMR. In fact, XMR
can be explained by the electron-hole compensation from
semiclassical two-band model\cite{sun2016large,PhysRevB.93.235142}.
In that case, the balance between electron concentration and hole
concentration will lead to unsaturated quadratic behavior of the MR,
and the value of MR depends on the mobility of carriers.

The topological property of the LnX family is interesting. A
previous theoretical work\cite{zeng2015topological} predicts that
LaX (X=N, P, As, Sb, Bi) are topological semimetals or topological
insulators. Later ARPES experiments show that LaSb is a
topologically trivial material without band
inversion\cite{PhysRevLett.117.127204} while LaBi is a topological
semimetal with multiple Dirac cones in the surface band
structure\cite{PhysRevB.94.165163,nayak2017multiple}. By drawing the
topological phase diagram of CeX (X=P, As, Sb, Bi) as a function of
the spin-orbit-coupling (SOC) effect, Kuroda \emph{et al.}
demonstrates the topological phase transition from trivial to
nontrivial with the increase of SOC
effect\cite{kuroda2017experimental}. Consequently, it is of interest
to explore the possible topological materials in other members of
LnX with strong SOC effect.

TmSb is an isostructural compound to LaSb/LaBi. In this work, we
have grown the high quality single crystals of TmSb and investigated
the detailed magneto-transport properties and the electronic
structure. The transverse MR of TmSb reaches 3.31$\times$10$^{4}\%$
at 2.3 K $\&$ 14 T. The split of Fermi surfaces (FSs) is found
through the analysis of SdH oscillation, which is attributed to
the nonsymmetric spin-orbit interaction. The angle-dependent MR are
measured to clarify the contribution of each Fermi surface (FS) to
the conductivity. In addition, the electronic structure of TmSb
has been studied by ARPES experiments and first-principles
calculations. The trivial Berry phase and the absence of band
inversion indicate that TmSb is a topologically trivial semimetal.
The Hall measurements reveal the compensation of carriers and the
high mobility, which constitute the origin of the observed XMR.

\section{Experimental methods and crystal structure}

\begin{figure}[htbp]
\centering
\includegraphics[width=0.48\textwidth]{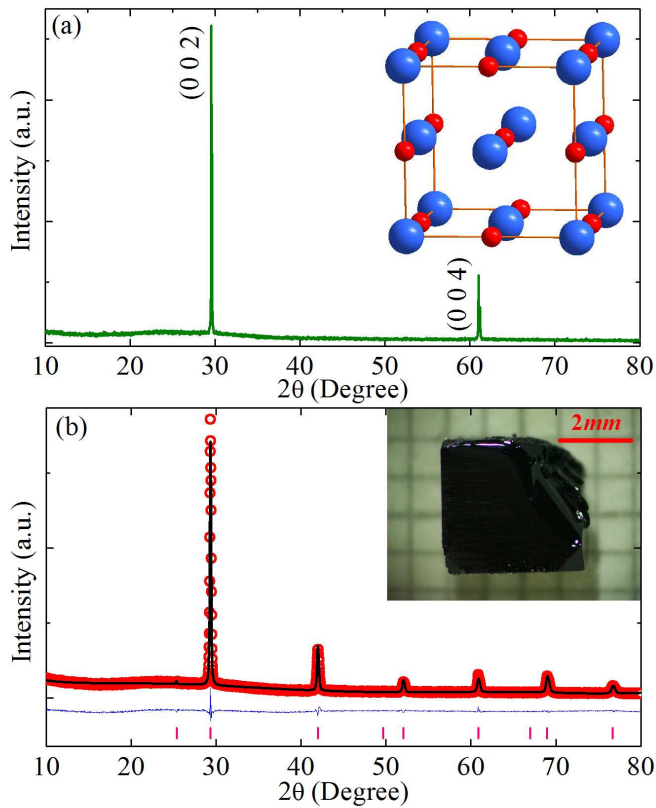}
\caption{(a) Single crystal XRD pattern of a TmSb crystal, showing
only the (\emph{00l}) reflections. Inset: the crystal structure of
TmSb. The blue and red balls represent Tm and Sb, respectively. (b)
Powder XRD pattern of TmSb with refinement. Red circle and black
solid line represent the data of experiment and the fit curve,
respectively. The difference plot is in blue. The pink vertical
lines denote the positions of Bragg peaks of TmSb. The inset is an
image of TmSb single crystal.}\end{figure}

Single crystals of TmSb were grown from Sb flux. Tm and excess Sb
were placed in a crucible with a ratio of Tm:Sb=1:6. Then the
crucible was sealed into an evacuated quartz tube and heated to
1150$^0$C. After cooling to 750$^0$C in 300 hours, the excess antimony
flux was removed with centrifuge. The elemental composition was
checked by energy dispersive x-ray spectroscopy (EDS, Oxford X-Max
50). X-ray diffraction (XRD) patterns of powder and single crystal
were collected from a Bruker D8 Advance x-ray diffractometer using
Cu K$_{\alpha}$ radiation. TOPAS-4.2 was employed for the
refinement. Resistivity measurements were performed on a Quantum
Design physical property measurement system (QD PPMS-14T). ARPES
measurements were taken at the Dreamline beamline of the Shanghai
Synchrotron Radiation Facility (SSRF). The crystals were cleaved \emph{in situ} along the (\emph{001}) plane and measured at T$\sim$20 K with a
working vacuum better than 5$\times$10$^{-11}$ Torr. The
first-principles calculations were performed with the projector
augmented wave (PAW) method\cite{PhysRevB.50.17953,PhysRevB.59.1758}
as implemented in the VASP
package\cite{PhysRevB.47.558,*kresse1996efficiency,*PhysRevB.54.11169}.
For the exchange-correlation functional, we adopted the generalized
gradient approximation (GGA) of Perdew-Burke-Ernzerhof (PBE)
type\cite{PhysRevLett.77.3865}. The kinetic energy cutoff of the
plane-wave basis was set to be 250 eV. The Brillouin zone was
sampled with a 20$\times$20$\times$20 $k$-point mesh and the
Gaussian smearing method with a width of 0.05 eV was used to broaden
the Fermi surface. Both cell parameters and internal atomic
positions were fully relaxed until all forces became less than 0.01
eV/${\AA}$. The calculated lattice constant 6.131 ${\AA}$ of TmSb
agrees well with the experimental value 6.105
${\AA}$\cite{abdusalyamova1994synthesis}. In the study of electronic
structure, the modified Becke-Johnson
(MBJ)\cite{becke2006simple,*PhysRevLett.102.226401} exchange
potential at the meta-GGA level of the Jacob¡¯s ladder was used and
the SOC effect was included. For the calculations of Fermi surfaces,
the maximally localized Wannier functions¡¯ (MLWF)
method\cite{PhysRevB.56.12847,PhysRevB.65.035109} was employed. TmSb
crystallizes in the NaCl-type structure as shown in the inset of
Fig. 1(a). The obtained TmSb crystals are in the shape of cubes. The single
crystal XRD pattern indicates that the surface of the crystal is the
(\emph{0 0 l}) plane (Fig. 1(a)). The powder XRD pattern of TmSb
crystals can be well refined as shown in Fig. 1(b). The refined
lattice parameter \emph{a} (6.08(0)${\AA}$) is in good agreement with the
value in Inorganic Crystal Structure Database
(ICSD)\cite{abdusalyamova1994synthesis}.
\section{Results and discussion}

\begin{figure*}[htbp]
\centering
\includegraphics[width=\textwidth]{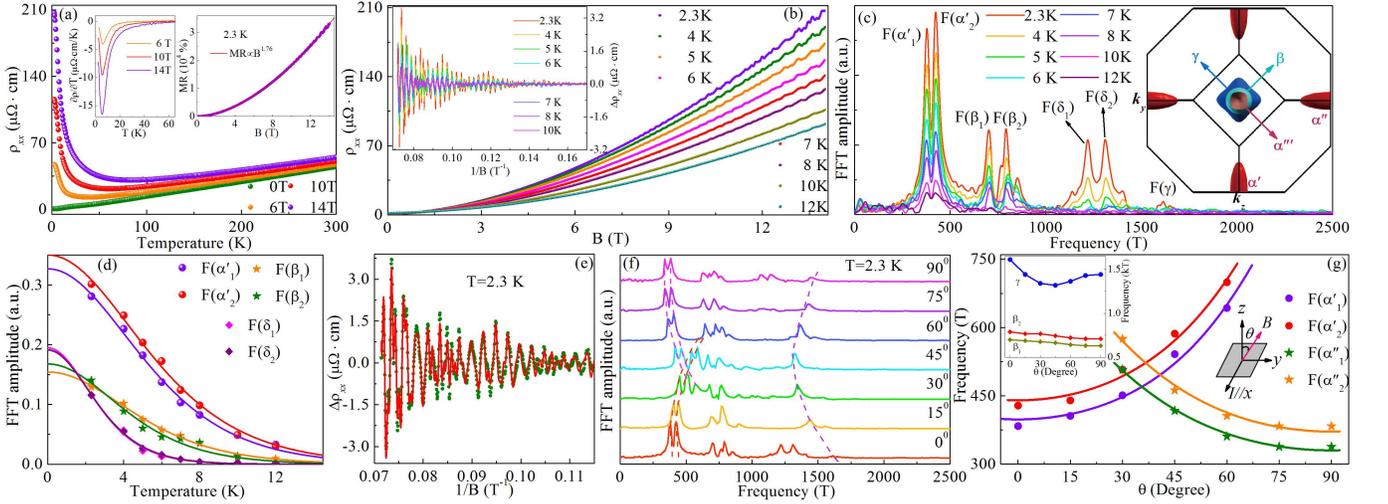}
\caption{Magneto-transport properties of TmSb(Sample 1, RRR=70). (a)
Temperature dependence of resistivity $\rho_{xx}(T)$ at B=0 T, 6 T,
10 T, 14 T. Inset on the left: $\partial \rho/\partial T$ as a
function of temperature. Inset on the right: MR versus magnetic
field B at 2.3 K. The MR follows B$^{1.76}$, which can be well
fitted as shown in the red solid line. (b) Magnetic field dependence
of resistivity $\rho_{xx}(B)$ at different temperatures. Inset: The
amplitude of SdH oscillation plotted as a function of 1/B. (c) The
FFT spectra of the corresponding oscillations. Inset: The projection
of the calculated FSs from the direction of $k_x$. (d) Temperature
dependent FFT amplitude of the frequencies. The solid lines are
fittings using the thermal factor in LK formula. (e) The fit (red
solid line) of SdH oscillation at 2.3 K using the multiband LK
formula. (f) FFT spectra of the SdH oscillations with the change of
$\theta$ at 2.3 K. (g) The frequencies originating from electron-like
FSs plotted as a function of the angle $\theta$. The solid lines are
fits to the equation presented in the text. The inset on the left
shows angle-dependence of the frequencies originating from hole-like
FSs. The inset on the right is a schematic diagram of the
measurements.}
\end{figure*}

We have investigated the magneto-transport properties of TmSb in
detail. Figure 2(a) shows the temperature dependent resistivity
$\rho_{xx}(T)$ under different magnetic fields. TmSb exhibits
metallic behavior under zero magnetic field. After applying a
moderate field, an upturn appears in $\rho_{xx}(T)$ curve with the
temperature decreased. The upturn can be enhanced by increasing
magnetic field. Similar behavior has also been observed in the
isostructural compounds
LaSb/LaBi/YSb\cite{tafti2015resistivity,sun2016large,PhysRevB.93.241106,Yu2017Magnetoresistance}
and other XMR materials (such as WTe$_2$\cite{ali2014large},
NbAs$_2$/TaAs$_2$\cite{PhysRevB.94.041103}). Especially, in WTe$_2$,
the upturn has been successfully explained by the following of
Kohler's rule in high quality samples with low charge carrier
density\cite{PhysRevB.92.180402}. Resistivity plateau is another
phenomenon usually observed in XMR materials. The resistivity
plateau seems to be absent in TmSb. However, as seen in the
$\partial \rho/\partial T$ curves (inset on the left of Fig. 2(a))
derived from the main panel, a minimum at $T_i$$\sim$5.6 K can be
obtained under different fields, indicating that the resistivity
plateau starts to emerge. The resistivity plateau is suggested to
originate from the temperature-insensitive resistivity at zero
field\cite{PhysRevB.96.125112,PhysRevB.93.235142}. The inset on the
right of Fig. 2(a) plots the transverse MR of TmSb as a function of
field. The MR follows B$^{1.76}$ (red solid line) and the value
reaches 3.31$\times$10$^{4}\%$ at 2.3 K $\&$ 14 T. Usually, in
semimetals with perfect electron-hole compensation, the MR will
exhibit quadratic behavior (MR$\propto$B$^2$) and not be saturated.
The index in TmSb deviates from 2, indicating that the electron and
hole in TmSb may be slightly imbalanced.

SdH oscillation has been observed at low temperature and high field
(Fig. 2(b)). The oscillation becomes weaker with the increase of
temperature. After subtracting a smooth background, the SdH
oscillation amplitude $\Delta\rho_{xx}=\rho_{xx}-<\rho_{xx}>$ can be
obtained as shown in the inset of Fig. 2(b). Figure 2(c) presents
the fast Fourier transform (FFT) analysis of the SdH oscillation.
Seven peaks (including three pairs of peaks and one single peak) are
identified from the FFT spectra. Since the Onsager relation
$F=(\phi_0/2\pi^2)A=(\hbar/2\pi e)A$ describes that the frequency
$F$ is proportional to the extremal cross-sectional area $A$ of FS
normal to the field, three pairs of peaks mean the FSs split under
the field. In fact, the split of FSs has also been observed in de
Haas-van Alphen (dHvA) type oscillation of
TmSb\cite{nimori1995haas}. TmSb is
paramagnetic\cite{PhysRevB.1.1211}, and the large magnetization is
contributed by the local magnetic moment of Tm$^{3+}$ ions which
develops with the application of magnetic
field\cite{nimori1995haas}. The spin degeneracy is lifted under the
field and the nonsymmetric spin-orbit interaction is
formed\cite{onuki2014chiral}, resulting in the split of FSs.

The inset of Fig. 2(c) shows the projection of the calculated FSs on
$k_y$-$k_z$ plane. In the current measurement ($I//x$, $B//z$),
there are three kinds of electron-like FSs ($\alpha^{\prime}$,
$\alpha^{\prime\prime}$ and $\alpha^{\prime\prime\prime}$) based on
the difference of extremal cross-sectional area. The other two
hole-like FSs are denoted as $\beta$ and $\gamma$, respectively.
However, only the frequencies from $\alpha^{\prime}$, $\beta$ and
$\gamma$ are observed in the FFT spectra (F($\delta_1$) and
F($\delta_2$) come from the mixture of the FSs $\alpha^{\prime}$ and
$\alpha^{\prime\prime}$, which will be discussed below). The absence
of the frequencies from $\alpha^{\prime\prime}$ and
$\alpha^{\prime\prime\prime}$ is understandable. Since the field is
parallel to $z$ axis, the extremal cross-sectional area of
$\alpha^{\prime\prime}$ is close to that of $\beta$. So the
frequencies from $\alpha^{\prime\prime}$ mix with that from
$\beta$ and can not be separated in the FFT spectra. Rotating the
field will change the extremal cross-sectional area of
$\alpha^{\prime\prime}$ and make its corresponding frequencies
appear, which has been proven by the angle-dependent MR (see below).
For the elliptical FS $\alpha^{\prime\prime\prime}$, a possible
explanation is that the mobility along the long axis is much smaller
than the mobility along the short axis. Such anisotropic mobility
has been derived from the quantitative analysis in YSb/LaSb, where
both the anisotropy and multiband nature are considered\cite{PhysRevB.96.075159,PhysRevB.96.125112}.

\begin{figure*}[htbp]
\centering
\includegraphics[width=\textwidth]{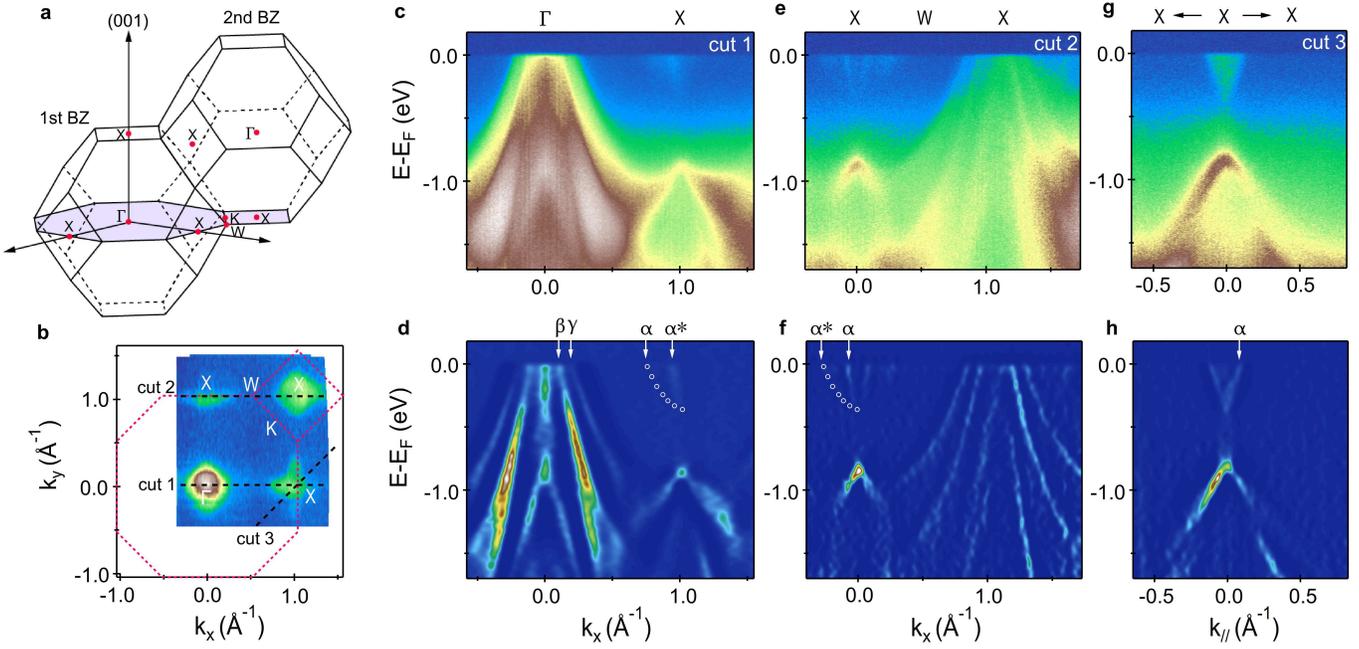}
\caption{Fermi surface intensity plot and band dispersions along
high-symmetry directions measured by ARPES. (a) Schematic of the
1$^{st}$ and 2$^{nd}$ 3D BZs with high symmetry points marked by red
points. The purple area illustrates the \emph{k}-space location of
the red lines in (b), which indicates the mapping area. (b) ARPES
intensity plot of TmSb close to $k_z$$\sim$0 with $hv$=53 eV at T$\sim$20 K with high symmetry points marked on it. (c), (e), (g) Photoemission intensity plots of cut1, cut2 and cut3 indicated in (b) respectively. (d), (f), (h) 2D curvature intensity
plots of (c), (e), (g) respectively, and white open circles in (d) and
(f) indicate half of the larger electron pockets at $X$
points.}\end{figure*}

The amplitude of SdH oscillation is described by Lifshitz-Kosevich (LK) formula:
\begin{equation}\label{equ1}
\centering
\Delta\rho\propto\frac{\lambda T}{sinh(\lambda T)}e^{-\lambda T_D}cos[2\pi(\frac{F}{B}-\frac{1}{2}+\beta+\delta)].
\end{equation}
In the formula, $\lambda= (2\pi^2k_{B}m^*)/(\hbar eB)$. $k_B$ and
$m^*$ are the Boltzmann constant and the effective mass of carrier,
respectively. $T_D$ is the Dingle temperature, and $2\pi \beta$ is
the Berry phase. $\delta$ is a phase shift, with the value of
$\delta=0$ and $\pm1/8$ for the 2D and 3D systems, respectively.
Figure 2(d) shows the temperature dependence of FFT amplitude of the
corresponding frequencies. The data can be well fitted by the
thermal factor $R_T=(\lambda T)/sinh(\lambda T)$ in LK formula. The
fitted effective masses (see Table I) are comparable with that of
LaSb\cite{tafti2016temperature} and NdSb\cite{PhysRevB.93.205152}.
As for F($\delta_1$) and F($\delta_2$), the effective masses are 0.554$m_e$ and 0.542$m_e$, respectively. Berry phase is a way to
roughly estimate the topological property of the materials. Since the
oscillation is multifrequency, we fit the oscillation pattern using
multiband LK formula (Fig. 2(e)) to obtain the values of Berry phase
and Dingle temperature. As shown in Table I, the values of Berry
phase are far away from the nontrivial value $\pi$, suggesting that
TmSb is possible topologically trivial material.

\begin{table}
  \centering
  \caption{Parameters derived from SdH oscillation. $F$, oscillation frequency; A, extremal cross-sectional area of FS normal to field; $k_F$, Fermi vector; $m^*$, effective mass; $T_D$, Dingle temperature; $2\pi\beta$, Berry phase.}
  \label{oscillations}
  \begin{tabular}{ccccccc}
    \hline\hline
    & $F$ (T) & A (${\AA}^{-2}$) & $k_F$ (${\AA}^{-1}$) & $m^*/m_e$ & $T_D$ (K) & $2\pi\beta$ \\
    \hline
$\alpha^{\prime}_{1}$ & 383.5 & 0.037 & 0.108 & 0.278 & 11.4 & 0.38$\pi$-0.25$\pi$ \\
$\alpha^{\prime}_{2}$ & 428.6 & 0.041 & 0.114 & 0.264 & 10.5 & -0.27$\pi$-0.25$\pi$ \\
$\beta_{1}$ & 699.3 & 0.067 & 0.146 & 0.300 & 8.5 & 0.29$\pi$+0.25$\pi$ \\
$\beta_{2}$ & 795.2 & 0.076 & 0.155 & 0.345 & 5.8 & 0.29$\pi$+0.25$\pi$ \\
    \hline\hline
  \end{tabular}
\end{table}

\begin{figure}[htbp]
\centering
\includegraphics[width=0.48\textwidth]{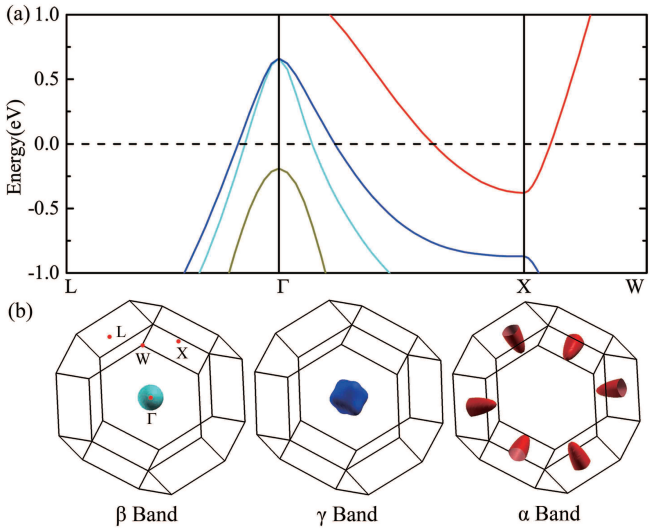}
\caption{(a) Band structure along high-symmetry directions of the Brillouin zone and (b) Fermi surfaces of TmSb calculated with the MBJ potential and including the SOC effect. The Fermi level is set to zero.}
\end{figure}

Angle-dependent MR measurements are performed to further understand
the contribution of each FS. Figure 2(f) shows the FFT spectra of
SdH oscillations with rotating the field in $y$-$z$ plane. With the
$\theta$ changing from 0$^0$ to 90$^0$, the extremal cross-sectional
area of $\alpha^{\prime}$ normal to field increases while that of
$\alpha^{\prime\prime}$ decreases. As a result, the frequencies from
$\alpha^{\prime}$ increases, and the frequencies from
$\alpha^{\prime\prime}$ can be identified when $\theta$=30$^0$
before decreasing with angle gradually. The angle-dependent
frequencies from $\beta$ are nearly unchanged while the frequency
from $\gamma$ varies slightly.

\begin{figure}[htbp]
\centering
\includegraphics[width=0.48\textwidth]{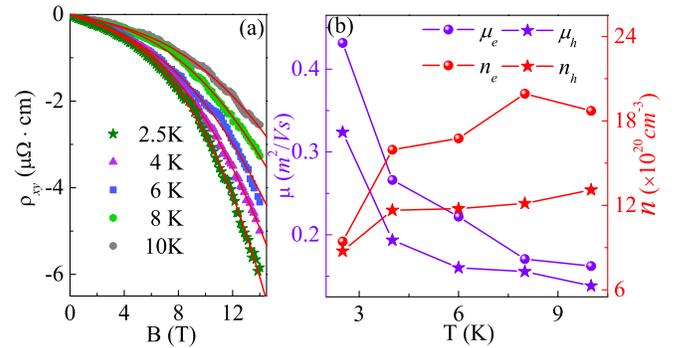}
\caption{(a) Magnetic field dependence of Hall resistivity at
different temperatures (Sample 2, RRR=24.1,
MR$_{2.8K,14T}$=7.62$\times$10$^{3}\%$). The red solid lines are the
fits using two-band model. (b) The obtained carriers¡¯
concentrations and mobility from the fits.}
\end{figure}

Figure 2(g) presents the angle-dependence of the frequencies.
Two-dimensional FS is suggested to exist in LaSb since the frequency
$F(\theta)$ follows
$F(0)/cos(\theta-n\pi/2)$\cite{tafti2015resistivity}. However, the
data in TmSb can't be well fitted (not presented here) by the above function. In fact, it is suggested to be a
pseudo-two-dimensional characteristic of ellipsoidal
FS\cite{PhysRevB.93.241106}, because the extremal cross-sectional
area $A=\pi ab/\sqrt{sin^2\theta+(a^2/b^2)cos^2\theta}$ (\emph{a}
and \emph{b} are the semimajor and semiminor axes of the ellipsoid,
respectively) can be approximated as $\pi b^2/cos\theta$ for small
$\theta$ values and $a$$\gg$$b$. Reasonably, the equation
$F(\theta)=F(0)/\sqrt{(b/a)^{2}sin^2(\theta-n\pi/2)+cos^2(\theta-n\pi/2)}$
($n$=0, 1 for $\alpha^{\prime}_{1}$($\alpha^{\prime}_{2}$),
$\alpha^{\prime\prime}_{1}$($\alpha^{\prime\prime}_{2}$),
respectively) is employed to describe the angle-dependent
frequencies and the experimental data can be well fitted as shown by
the solid lines in Fig. 2(g). The obtained $a/b$ of FS
$\alpha^{\prime\prime}_{1}$ ($\alpha^{\prime\prime}_{2}$) is 2.06
(2.07). Then the values of F($\alpha^{\prime\prime}_{1}$)=697.0 T
and F($\alpha^{\prime\prime}_{2}$)=793.8 T at $\theta$=0$^0$ can be
derived, which are close to F($\beta_1$) and F($\beta_2$) as
expected. Then the frequency F($\delta_1$) can be identified as
F($\alpha^{\prime}_{2}$)+F($\alpha^{\prime\prime}_{2}$), implying
the merging of the corresponding FSs. F($\delta_2$) is the split
frequency of F($\delta_1$) under the field, which has a similar
effective mass as F($\delta_1$). As shown in the inset of Fig. 2(g),
with the change of $\theta$, the frequencies from $\beta$ are nearly
unchanged while the frequency from $\gamma$ varies slightly. Such
behaviors are expected since the FS $\beta$ is nearly spherical and
the FS $\gamma$ is slightly anisotropic. The behavior of
angle-dependent frequencies is clearly related to the shape of FSs.

ARPES measurements were performed to reveal the electronic structure of TmSb. TmSb crystalizes in a face-centered cubic (FCC) structure. The first and second three-dimensional Brillouin zones (BZs) are shown in Fig. 3(a). ARPES measurements were performed at T$\sim$20 K at a photon energy of 53 eV. Figure 3(b) shows the measured Fermi surface map close to the $k_z$$\sim$0 plane, which contains pockets at the $\Gamma$ and $X$ points. To reveal their dispersions, we show in Figs. 3(c), 3(e) and 3(g) cuts through these pockets as indicated by lines in Fig. 3(b). To enhance the dispersing bands, the corresponding curvature\cite{zhang2011precise} plots are also shown in Figs. 3(d), 3(f) and 3(h). The pockets at the $\Gamma$ point are clearly identified to be hole pockets and labeled by $\beta$ and $\gamma$  in Figs. 3(d). The ellipsoid-like pocket at each $X$ point is labeled by $\alpha$ in Figs. 3(d), 3(f) and 3(h), which is also seen in calculations in Fig. 4(b) with its long axis along the $\Gamma$-$X$ direction. The existence of another electron pocket $\alpha^*$ in Figs. 3(e) and 3(f) is likely caused by the pocket at the $k_z$$\sim$0.5 plane (top or bottom of the Brillouin zone) due to the $k_z$ broadening. In all the cuts, an energy gap of $\sim$ 0.5 eV is observed at the $X$ point between the conduction band $\alpha$ and the valence bands. The absence of band anti-crossing along the $\Gamma$-$X$ direction indicates the topologically trivial characteristic of TmSb, which is similar to the case of LaSb and YSb\cite{PhysRevLett.117.127204,PhysRevLett.117.267201,Yu2017Magnetoresistance}.

First-principles calculations have also been employed to study the
electronic structure of TmSb. As shown in Fig. 4(a), the calculated
band structure is quite consistent with that observed by ARPES.
There are two hole bands ($\beta$ and $\gamma$) and one electron
band ($\alpha$) crossing the Fermi level. The gap at $X$ point is
about 0.49 eV. Combined with the trivial Berry phase obtained from
SdH oscillation and the electronic structure revealed by ARPES
experiments and first-principles calculations, TmSb is suggested to
be a topologically trivial semimetal. Figure. 4(b) presents the
calculated FSs of TmSb with the SOC effect included. The colors of
the FSs are in a one-to-one relationship with the corresponding
bands crossing the Fermi level. For the two hole pockets, $\beta$ is
nearly spherical, but $\gamma$ has a FS stretched in the $\{100\}$
directions. The electron pockets $\alpha$ are ellipsoidal and
located at every $X$ point.

Since the topological trivial characteristic of TmSb has been
confirmed in the above discussion, the breaking of topological
protection is not suitable to explain the origin of XMR in TmSb.
Hall measurements are taken to achieve the information about
carriers and to reveal the origin of XMR in TmSb. Figure 5(a) shows
the field dependence of Hall resistivity
$\rho_{xy}=[\rho_{xy}(+B)-\rho_{xy}(-B)]/2$ of TmSb. The $\rho_{xy}$
curves are nonlinear, indicating that the electron and hole coexist
in TmSb. The Hall resistivity can be described by the semiclassical
two-band model:
\begin{equation}\label{equ2}
\rho_{xy}=\frac{B}{e}\frac{(n_h \mu_h^2-n_e \mu_e^2)+(n_h-n_e)(\mu_h \mu_e)^2 B^2}{(n_h \mu_h+n_e \mu_e)^2+(n_h-n_e)^2 (\mu_h \mu_e)^2 B^2},
\end{equation}
where $\mu_h(\mu_e)$ and $n_h(n_e)$ are hole (electron) mobility and
hole (electron) concentration, respectively. As shown by the red
solid lines in Fig. 5(a), the data can be well fitted by the
two-band model. Figure 5(b) presents the temperature dependent
carriers' concentrations and mobility, which are derived from the
fitting. The concentrations at 2.5 K are
$n_e=9.43\times10^{20}cm^{-3}$ and $n_h=8.75\times10^{20}cm^{-3}$.
The ratio $n_h/n_e\approx0.93$ indicates the compensation of hole
and electron in TmSb. The values of mobility at 2.5 K
($\mu_e=4.32\times10^3cm^2V^{-1}s^{-1}$,
$\mu_h=3.24\times10^3cm^2V^{-1}s^{-1}$) are lower than those of
LaSb/LaBi/YSb\cite{tafti2016temperature,sun2016large,Yu2017Magnetoresistance},
which may be caused by the lower quality of the sample used to
measure the Hall resistivity. In the case of electron-hole
compensation ($n_e$=$n_h$), the relation MR=$\mu_e\mu_h$B$^2$ can be
derived from the field dependent resistivity:
\begin{equation}\label{equ3}
\rho (B)=\frac{(n_h \mu_h+n_e \mu_e)+(n_h \mu_e+n_e \mu_h)\mu_h \mu_e B^2}{e (n_h \mu_h+n_e \mu_e)^2+e (n_h-n_e)^2 (\mu_h \mu_e)^2 B^2}.
\end{equation}
Obviously, the MR will be unsaturated, and the value of MR depends
on the mobility of carriers. The compensated carrier concentrations
and high mobility are suggested to be responsible for the XMR in
TmSb.

\section{Summary}

In summary, single crystals of TmSb are grown and the
magneto-transport properties have been investigated. Analysis on the
FFT spectra of the SdH oscillations observed at low temperature and
high field clearly indicates the splits of Fermi surfaces. The
extracted trivial Berry phase from the fit of LK formula, combining
with the electronic structures from ARPES measurements and
first-principles calculations confirm that TmSb is a semimetal with
topologically trivial band structures and nearly compensated
concentrations of electron and hole. The XMR in TmSb is attributed
to the electron-hole compensation and high mobility of carriers.

\section{Acknowledgments}

We thank Peng-Jie Guo for helpful discussions. This work is supported by the National Natural Science Foundation of China (No.11574391, No.11774424, No.11474356, and No.91421304), the Fundamental Research Funds for the Central Universities, and the Research Funds of Renmin University of China (No.14XNLQ07, No.14XNLQ03, and No.16XNLQ01). Computational resources have been provided by the Physical Laboratory of High Performance Computing at Renmin University of China. The Fermi surfaces were prepared with the XCRYSDEN program\cite{kokalj2003computer}.

\bibliography{Bibtex}

\begin{thebibliography}{51}%
\makeatletter
\providecommand \@ifxundefined [1]{%
 \@ifx{#1\undefined}
}%
\providecommand \@ifnum [1]{%
 \ifnum #1\expandafter \@firstoftwo
 \else \expandafter \@secondoftwo
 \fi
}%
\providecommand \@ifx [1]{%
 \ifx #1\expandafter \@firstoftwo
 \else \expandafter \@secondoftwo
 \fi
}%
\providecommand \natexlab [1]{#1}%
\providecommand \enquote  [1]{``#1''}%
\providecommand \bibnamefont  [1]{#1}%
\providecommand \bibfnamefont [1]{#1}%
\providecommand \citenamefont [1]{#1}%
\providecommand \href@noop [0]{\@secondoftwo}%
\providecommand \href [0]{\begingroup \@sanitize@url \@href}%
\providecommand \@href[1]{\@@startlink{#1}\@@href}%
\providecommand \@@href[1]{\endgroup#1\@@endlink}%
\providecommand \@sanitize@url [0]{\catcode `\\12\catcode `\$12\catcode
  `\&12\catcode `\#12\catcode `\^12\catcode `\_12\catcode `\%12\relax}%
\providecommand \@@startlink[1]{}%
\providecommand \@@endlink[0]{}%
\providecommand \url  [0]{\begingroup\@sanitize@url \@url }%
\providecommand \@url [1]{\endgroup\@href {#1}{\urlprefix }}%
\providecommand \urlprefix  [0]{URL }%
\providecommand \Eprint [0]{\href }%
\providecommand \doibase [0]{http://dx.doi.org/}%
\providecommand \selectlanguage [0]{\@gobble}%
\providecommand \bibinfo  [0]{\@secondoftwo}%
\providecommand \bibfield  [0]{\@secondoftwo}%
\providecommand \translation [1]{[#1]}%
\providecommand \BibitemOpen [0]{}%
\providecommand \bibitemStop [0]{}%
\providecommand \bibitemNoStop [0]{.\EOS\space}%
\providecommand \EOS [0]{\spacefactor3000\relax}%
\providecommand \BibitemShut  [1]{\csname bibitem#1\endcsname}%
\let\auto@bib@innerbib\@empty
%</preamble>
\bibitem [{\citenamefont {Zeng}\ \emph {et~al.}(2015)\citenamefont {Zeng},
  \citenamefont {Fang}, \citenamefont {Chang}, \citenamefont {Chen},
  \citenamefont {Hsieh}, \citenamefont {Bansil}, \citenamefont {Lin},\ and\
  \citenamefont {Fu}}]{zeng2015topological}%
  \BibitemOpen
  \bibfield  {author} {\bibinfo {author} {\bibfnamefont {M.}~\bibnamefont
  {Zeng}}, \bibinfo {author} {\bibfnamefont {C.}~\bibnamefont {Fang}}, \bibinfo
  {author} {\bibfnamefont {G.}~\bibnamefont {Chang}}, \bibinfo {author}
  {\bibfnamefont {Y.-A.}\ \bibnamefont {Chen}}, \bibinfo {author}
  {\bibfnamefont {T.}~\bibnamefont {Hsieh}}, \bibinfo {author} {\bibfnamefont
  {A.}~\bibnamefont {Bansil}}, \bibinfo {author} {\bibfnamefont
  {H.}~\bibnamefont {Lin}}, \ and\ \bibinfo {author} {\bibfnamefont
  {L.}~\bibnamefont {Fu}},\ }\href@noop {} {\bibfield  {journal} {\bibinfo
  {journal} {arXiv preprint arXiv:1504.03492}\ } (\bibinfo {year}
  {2015})}\BibitemShut {NoStop}%
\bibitem [{\citenamefont {Tafti}\ \emph
  {et~al.}(2016{\natexlab{a}})\citenamefont {Tafti}, \citenamefont {Gibson},
  \citenamefont {Kushwaha}, \citenamefont {Haldolaarachchige},\ and\
  \citenamefont {Cava}}]{tafti2015resistivity}%
  \BibitemOpen
  \bibfield  {author} {\bibinfo {author} {\bibfnamefont {F.~F.}\ \bibnamefont
  {Tafti}}, \bibinfo {author} {\bibfnamefont {Q.~D.}\ \bibnamefont {Gibson}},
  \bibinfo {author} {\bibfnamefont {S.~K.}\ \bibnamefont {Kushwaha}}, \bibinfo
  {author} {\bibfnamefont {N.}~\bibnamefont {Haldolaarachchige}}, \ and\
  \bibinfo {author} {\bibfnamefont {R.~J.}\ \bibnamefont {Cava}},\ }\href@noop
  {} {\bibfield  {journal} {\bibinfo  {journal} {Nat. Phys.}\ }\textbf
  {\bibinfo {volume} {12}},\ \bibinfo {pages} {272} (\bibinfo {year}
  {2016}{\natexlab{a}})}\BibitemShut {NoStop}%
\bibitem [{\citenamefont {Tafti}\ \emph
  {et~al.}(2016{\natexlab{b}})\citenamefont {Tafti}, \citenamefont {Gibson},
  \citenamefont {Kushwaha}, \citenamefont {Krizan}, \citenamefont
  {Haldolaarachchige},\ and\ \citenamefont {Cava}}]{tafti2016temperature}%
  \BibitemOpen
  \bibfield  {author} {\bibinfo {author} {\bibfnamefont {F.~F.}\ \bibnamefont
  {Tafti}}, \bibinfo {author} {\bibfnamefont {Q.}~\bibnamefont {Gibson}},
  \bibinfo {author} {\bibfnamefont {S.}~\bibnamefont {Kushwaha}}, \bibinfo
  {author} {\bibfnamefont {J.~W.}\ \bibnamefont {Krizan}}, \bibinfo {author}
  {\bibfnamefont {N.}~\bibnamefont {Haldolaarachchige}}, \ and\ \bibinfo
  {author} {\bibfnamefont {R.~J.}\ \bibnamefont {Cava}},\ }\href@noop {}
  {\bibfield  {journal} {\bibinfo  {journal} {Proc. Natl. Acad. Sci. U. S. A.}\
  }\textbf {\bibinfo {volume} {113}},\ \bibinfo {pages} {E3475} (\bibinfo
  {year} {2016}{\natexlab{b}})}\BibitemShut {NoStop}%
\bibitem [{\citenamefont {Zeng}\ \emph {et~al.}(2016)\citenamefont {Zeng},
  \citenamefont {Lou}, \citenamefont {Wu}, \citenamefont {Xu}, \citenamefont
  {Guo}, \citenamefont {Kong}, \citenamefont {Zhong}, \citenamefont {Ma},
  \citenamefont {Fu}, \citenamefont {Richard}, \citenamefont {Wang},
  \citenamefont {Liu}, \citenamefont {Lu}, \citenamefont {Huang}, \citenamefont
  {Fang}, \citenamefont {Sun}, \citenamefont {Wang}, \citenamefont {Wang},
  \citenamefont {Shi}, \citenamefont {Weng}, \citenamefont {Lei}, \citenamefont
  {Liu}, \citenamefont {Wang}, \citenamefont {Qian}, \citenamefont {Luo},\ and\
  \citenamefont {Ding}}]{PhysRevLett.117.127204}%
  \BibitemOpen
  \bibfield  {author} {\bibinfo {author} {\bibfnamefont {L.-K.}\ \bibnamefont
  {Zeng}}, \bibinfo {author} {\bibfnamefont {R.}~\bibnamefont {Lou}}, \bibinfo
  {author} {\bibfnamefont {D.-S.}\ \bibnamefont {Wu}}, \bibinfo {author}
  {\bibfnamefont {Q.~N.}\ \bibnamefont {Xu}}, \bibinfo {author} {\bibfnamefont
  {P.-J.}\ \bibnamefont {Guo}}, \bibinfo {author} {\bibfnamefont {L.-Y.}\
  \bibnamefont {Kong}}, \bibinfo {author} {\bibfnamefont {Y.-G.}\ \bibnamefont
  {Zhong}}, \bibinfo {author} {\bibfnamefont {J.-Z.}\ \bibnamefont {Ma}},
  \bibinfo {author} {\bibfnamefont {B.-B.}\ \bibnamefont {Fu}}, \bibinfo
  {author} {\bibfnamefont {P.}~\bibnamefont {Richard}}, \bibinfo {author}
  {\bibfnamefont {P.}~\bibnamefont {Wang}}, \bibinfo {author} {\bibfnamefont
  {G.~T.}\ \bibnamefont {Liu}}, \bibinfo {author} {\bibfnamefont
  {L.}~\bibnamefont {Lu}}, \bibinfo {author} {\bibfnamefont {Y.-B.}\
  \bibnamefont {Huang}}, \bibinfo {author} {\bibfnamefont {C.}~\bibnamefont
  {Fang}}, \bibinfo {author} {\bibfnamefont {S.-S.}\ \bibnamefont {Sun}},
  \bibinfo {author} {\bibfnamefont {Q.}~\bibnamefont {Wang}}, \bibinfo {author}
  {\bibfnamefont {L.}~\bibnamefont {Wang}}, \bibinfo {author} {\bibfnamefont
  {Y.-G.}\ \bibnamefont {Shi}}, \bibinfo {author} {\bibfnamefont {H.~M.}\
  \bibnamefont {Weng}}, \bibinfo {author} {\bibfnamefont {H.-C.}\ \bibnamefont
  {Lei}}, \bibinfo {author} {\bibfnamefont {K.}~\bibnamefont {Liu}}, \bibinfo
  {author} {\bibfnamefont {S.-C.}\ \bibnamefont {Wang}}, \bibinfo {author}
  {\bibfnamefont {T.}~\bibnamefont {Qian}}, \bibinfo {author} {\bibfnamefont
  {J.-L.}\ \bibnamefont {Luo}}, \ and\ \bibinfo {author} {\bibfnamefont
  {H.}~\bibnamefont {Ding}},\ }\href {\doibase 10.1103/PhysRevLett.117.127204}
  {\bibfield  {journal} {\bibinfo  {journal} {Phys. Rev. Lett.}\ }\textbf
  {\bibinfo {volume} {117}},\ \bibinfo {pages} {127204} (\bibinfo {year}
  {2016})}\BibitemShut {NoStop}%
\bibitem [{\citenamefont {Ban}\ \emph {et~al.}(2017)\citenamefont {Ban},
  \citenamefont {Guo}, \citenamefont {Luo},\ and\ \citenamefont
  {Wang}}]{Ban2017observation}%
  \BibitemOpen
  \bibfield  {author} {\bibinfo {author} {\bibfnamefont {W.-J.}\ \bibnamefont
  {Ban}}, \bibinfo {author} {\bibfnamefont {W.-T.}\ \bibnamefont {Guo}},
  \bibinfo {author} {\bibfnamefont {J.-L.}\ \bibnamefont {Luo}}, \ and\
  \bibinfo {author} {\bibfnamefont {N.-L.}\ \bibnamefont {Wang}},\ }\href@noop
  {} {\bibfield  {journal} {\bibinfo  {journal} {Chin. Phys. Lett.}\ }\textbf
  {\bibinfo {volume} {34}},\ \bibinfo {pages} {077804} (\bibinfo {year}
  {2017})}\BibitemShut {NoStop}%
\bibitem [{\citenamefont {Guo}\ \emph {et~al.}(2017)\citenamefont {Guo},
  \citenamefont {Yang}, \citenamefont {Liu},\ and\ \citenamefont
  {Lu}}]{PhysRevB.96.081112}%
  \BibitemOpen
  \bibfield  {author} {\bibinfo {author} {\bibfnamefont {P.-J.}\ \bibnamefont
  {Guo}}, \bibinfo {author} {\bibfnamefont {H.-C.}\ \bibnamefont {Yang}},
  \bibinfo {author} {\bibfnamefont {K.}~\bibnamefont {Liu}}, \ and\ \bibinfo
  {author} {\bibfnamefont {Z.-Y.}\ \bibnamefont {Lu}},\ }\href {\doibase
  10.1103/PhysRevB.96.081112} {\bibfield  {journal} {\bibinfo  {journal} {Phys.
  Rev. B}\ }\textbf {\bibinfo {volume} {96}},\ \bibinfo {pages} {081112}
  (\bibinfo {year} {2017})}\BibitemShut {NoStop}%
\bibitem [{\citenamefont {Han}\ \emph {et~al.}(2017)\citenamefont {Han},
  \citenamefont {Xu}, \citenamefont {Botana}, \citenamefont {Xiao},
  \citenamefont {Wang}, \citenamefont {Yang}, \citenamefont {Chung},
  \citenamefont {Kanatzidis}, \citenamefont {Norman}, \citenamefont
  {Crabtree},\ and\ \citenamefont {Kwok}}]{PhysRevB.96.125112}%
  \BibitemOpen
  \bibfield  {author} {\bibinfo {author} {\bibfnamefont {F.}~\bibnamefont
  {Han}}, \bibinfo {author} {\bibfnamefont {J.}~\bibnamefont {Xu}}, \bibinfo
  {author} {\bibfnamefont {A.~S.}\ \bibnamefont {Botana}}, \bibinfo {author}
  {\bibfnamefont {Z.~L.}\ \bibnamefont {Xiao}}, \bibinfo {author}
  {\bibfnamefont {Y.~L.}\ \bibnamefont {Wang}}, \bibinfo {author}
  {\bibfnamefont {W.~G.}\ \bibnamefont {Yang}}, \bibinfo {author}
  {\bibfnamefont {D.~Y.}\ \bibnamefont {Chung}}, \bibinfo {author}
  {\bibfnamefont {M.~G.}\ \bibnamefont {Kanatzidis}}, \bibinfo {author}
  {\bibfnamefont {M.~R.}\ \bibnamefont {Norman}}, \bibinfo {author}
  {\bibfnamefont {G.~W.}\ \bibnamefont {Crabtree}}, \ and\ \bibinfo {author}
  {\bibfnamefont {W.~K.}\ \bibnamefont {Kwok}},\ }\href {\doibase
  10.1103/PhysRevB.96.125112} {\bibfield  {journal} {\bibinfo  {journal} {Phys.
  Rev. B}\ }\textbf {\bibinfo {volume} {96}},\ \bibinfo {pages} {125112}
  (\bibinfo {year} {2017})}\BibitemShut {NoStop}%
\bibitem [{\citenamefont {Sun}\ \emph {et~al.}(2016)\citenamefont {Sun},
  \citenamefont {Wang}, \citenamefont {Guo}, \citenamefont {Liu},\ and\
  \citenamefont {Lei}}]{sun2016large}%
  \BibitemOpen
  \bibfield  {author} {\bibinfo {author} {\bibfnamefont {S.}~\bibnamefont
  {Sun}}, \bibinfo {author} {\bibfnamefont {Q.}~\bibnamefont {Wang}}, \bibinfo
  {author} {\bibfnamefont {P.-J.}\ \bibnamefont {Guo}}, \bibinfo {author}
  {\bibfnamefont {K.}~\bibnamefont {Liu}}, \ and\ \bibinfo {author}
  {\bibfnamefont {H.}~\bibnamefont {Lei}},\ }\href@noop {} {\bibfield
  {journal} {\bibinfo  {journal} {New J. Phys.}\ }\textbf {\bibinfo {volume}
  {18}},\ \bibinfo {pages} {082002} (\bibinfo {year} {2016})}\BibitemShut
  {NoStop}%
\bibitem [{\citenamefont {Kumar}\ \emph {et~al.}(2016)\citenamefont {Kumar},
  \citenamefont {Shekhar}, \citenamefont {Wu}, \citenamefont {Leermakers},
  \citenamefont {Young}, \citenamefont {Zeitler}, \citenamefont {Yan},\ and\
  \citenamefont {Felser}}]{PhysRevB.93.241106}%
  \BibitemOpen
  \bibfield  {author} {\bibinfo {author} {\bibfnamefont {N.}~\bibnamefont
  {Kumar}}, \bibinfo {author} {\bibfnamefont {C.}~\bibnamefont {Shekhar}},
  \bibinfo {author} {\bibfnamefont {S.-C.}\ \bibnamefont {Wu}}, \bibinfo
  {author} {\bibfnamefont {I.}~\bibnamefont {Leermakers}}, \bibinfo {author}
  {\bibfnamefont {O.}~\bibnamefont {Young}}, \bibinfo {author} {\bibfnamefont
  {U.}~\bibnamefont {Zeitler}}, \bibinfo {author} {\bibfnamefont
  {B.}~\bibnamefont {Yan}}, \ and\ \bibinfo {author} {\bibfnamefont
  {C.}~\bibnamefont {Felser}},\ }\href {\doibase 10.1103/PhysRevB.93.241106}
  {\bibfield  {journal} {\bibinfo  {journal} {Phys. Rev. B}\ }\textbf {\bibinfo
  {volume} {93}},\ \bibinfo {pages} {241106} (\bibinfo {year}
  {2016})}\BibitemShut {NoStop}%
\bibitem [{\citenamefont {Wu}\ \emph {et~al.}(2016{\natexlab{a}})\citenamefont
  {Wu}, \citenamefont {Kong}, \citenamefont {Wang}, \citenamefont {Johnson},
  \citenamefont {Mou}, \citenamefont {Huang}, \citenamefont {Schrunk},
  \citenamefont {Bud'ko}, \citenamefont {Canfield},\ and\ \citenamefont
  {Kaminski}}]{PhysRevB.94.081108}%
  \BibitemOpen
  \bibfield  {author} {\bibinfo {author} {\bibfnamefont {Y.}~\bibnamefont
  {Wu}}, \bibinfo {author} {\bibfnamefont {T.}~\bibnamefont {Kong}}, \bibinfo
  {author} {\bibfnamefont {L.-L.}\ \bibnamefont {Wang}}, \bibinfo {author}
  {\bibfnamefont {D.~D.}\ \bibnamefont {Johnson}}, \bibinfo {author}
  {\bibfnamefont {D.}~\bibnamefont {Mou}}, \bibinfo {author} {\bibfnamefont
  {L.}~\bibnamefont {Huang}}, \bibinfo {author} {\bibfnamefont
  {B.}~\bibnamefont {Schrunk}}, \bibinfo {author} {\bibfnamefont {S.~L.}\
  \bibnamefont {Bud'ko}}, \bibinfo {author} {\bibfnamefont {P.~C.}\
  \bibnamefont {Canfield}}, \ and\ \bibinfo {author} {\bibfnamefont
  {A.}~\bibnamefont {Kaminski}},\ }\href {\doibase 10.1103/PhysRevB.94.081108}
  {\bibfield  {journal} {\bibinfo  {journal} {Phys. Rev. B}\ }\textbf {\bibinfo
  {volume} {94}},\ \bibinfo {pages} {081108} (\bibinfo {year}
  {2016}{\natexlab{a}})}\BibitemShut {NoStop}%
\bibitem [{\citenamefont {Lou}\ \emph {et~al.}(2017)\citenamefont {Lou},
  \citenamefont {Fu}, \citenamefont {Xu}, \citenamefont {Guo}, \citenamefont
  {Kong}, \citenamefont {Zeng}, \citenamefont {Ma}, \citenamefont {Richard},
  \citenamefont {Fang}, \citenamefont {Huang}, \citenamefont {Sun},
  \citenamefont {Wang}, \citenamefont {Wang}, \citenamefont {Shi},
  \citenamefont {Lei}, \citenamefont {Liu}, \citenamefont {Weng}, \citenamefont
  {Qian}, \citenamefont {Ding},\ and\ \citenamefont
  {Wang}}]{PhysRevB.95.115140}%
  \BibitemOpen
  \bibfield  {author} {\bibinfo {author} {\bibfnamefont {R.}~\bibnamefont
  {Lou}}, \bibinfo {author} {\bibfnamefont {B.-B.}\ \bibnamefont {Fu}},
  \bibinfo {author} {\bibfnamefont {Q.~N.}\ \bibnamefont {Xu}}, \bibinfo
  {author} {\bibfnamefont {P.-J.}\ \bibnamefont {Guo}}, \bibinfo {author}
  {\bibfnamefont {L.-Y.}\ \bibnamefont {Kong}}, \bibinfo {author}
  {\bibfnamefont {L.-K.}\ \bibnamefont {Zeng}}, \bibinfo {author}
  {\bibfnamefont {J.-Z.}\ \bibnamefont {Ma}}, \bibinfo {author} {\bibfnamefont
  {P.}~\bibnamefont {Richard}}, \bibinfo {author} {\bibfnamefont
  {C.}~\bibnamefont {Fang}}, \bibinfo {author} {\bibfnamefont {Y.-B.}\
  \bibnamefont {Huang}}, \bibinfo {author} {\bibfnamefont {S.-S.}\ \bibnamefont
  {Sun}}, \bibinfo {author} {\bibfnamefont {Q.}~\bibnamefont {Wang}}, \bibinfo
  {author} {\bibfnamefont {L.}~\bibnamefont {Wang}}, \bibinfo {author}
  {\bibfnamefont {Y.-G.}\ \bibnamefont {Shi}}, \bibinfo {author} {\bibfnamefont
  {H.~C.}\ \bibnamefont {Lei}}, \bibinfo {author} {\bibfnamefont
  {K.}~\bibnamefont {Liu}}, \bibinfo {author} {\bibfnamefont {H.~M.}\
  \bibnamefont {Weng}}, \bibinfo {author} {\bibfnamefont {T.}~\bibnamefont
  {Qian}}, \bibinfo {author} {\bibfnamefont {H.}~\bibnamefont {Ding}}, \ and\
  \bibinfo {author} {\bibfnamefont {S.-C.}\ \bibnamefont {Wang}},\ }\href
  {\doibase 10.1103/PhysRevB.95.115140} {\bibfield  {journal} {\bibinfo
  {journal} {Phys. Rev. B}\ }\textbf {\bibinfo {volume} {95}},\ \bibinfo
  {pages} {115140} (\bibinfo {year} {2017})}\BibitemShut {NoStop}%
\bibitem [{\citenamefont {Nayak}\ \emph {et~al.}(2017)\citenamefont {Nayak},
  \citenamefont {Wu}, \citenamefont {Kumar}, \citenamefont {Shekhar},
  \citenamefont {Singh}, \citenamefont {Fink}, \citenamefont {Rienks},
  \citenamefont {Fecher}, \citenamefont {Parkin}, \citenamefont {Yan} \emph
  {et~al.}}]{nayak2017multiple}%
  \BibitemOpen
  \bibfield  {author} {\bibinfo {author} {\bibfnamefont {J.}~\bibnamefont
  {Nayak}}, \bibinfo {author} {\bibfnamefont {S.-C.}\ \bibnamefont {Wu}},
  \bibinfo {author} {\bibfnamefont {N.}~\bibnamefont {Kumar}}, \bibinfo
  {author} {\bibfnamefont {C.}~\bibnamefont {Shekhar}}, \bibinfo {author}
  {\bibfnamefont {S.}~\bibnamefont {Singh}}, \bibinfo {author} {\bibfnamefont
  {J.}~\bibnamefont {Fink}}, \bibinfo {author} {\bibfnamefont {E.~E.}\
  \bibnamefont {Rienks}}, \bibinfo {author} {\bibfnamefont {G.~H.}\
  \bibnamefont {Fecher}}, \bibinfo {author} {\bibfnamefont {S.~S.}\
  \bibnamefont {Parkin}}, \bibinfo {author} {\bibfnamefont {B.}~\bibnamefont
  {Yan}},  \emph {et~al.},\ }\href@noop {} {\bibfield  {journal} {\bibinfo
  {journal} {Nat. Commun.}\ }\textbf {\bibinfo {volume} {8}},\ \bibinfo {pages}
  {13942} (\bibinfo {year} {2017})}\BibitemShut {NoStop}%
\bibitem [{\citenamefont {Singha}\ \emph {et~al.}(2017)\citenamefont {Singha},
  \citenamefont {Satpati},\ and\ \citenamefont {Mandal}}]{Singha2017Fermi}%
  \BibitemOpen
  \bibfield  {author} {\bibinfo {author} {\bibfnamefont {R.}~\bibnamefont
  {Singha}}, \bibinfo {author} {\bibfnamefont {B.}~\bibnamefont {Satpati}}, \
  and\ \bibinfo {author} {\bibfnamefont {P.}~\bibnamefont {Mandal}},\
  }\href@noop {} {\bibfield  {journal} {\bibinfo  {journal} {Sci Rep}\ }\textbf
  {\bibinfo {volume} {7}},\ \bibinfo {pages} {6321} (\bibinfo {year}
  {2017})}\BibitemShut {NoStop}%
\bibitem [{\citenamefont {Guo}\ \emph {et~al.}(2016)\citenamefont {Guo},
  \citenamefont {Yang}, \citenamefont {Zhang}, \citenamefont {Liu},\ and\
  \citenamefont {Lu}}]{PhysRevB.93.235142}%
  \BibitemOpen
  \bibfield  {author} {\bibinfo {author} {\bibfnamefont {P.-J.}\ \bibnamefont
  {Guo}}, \bibinfo {author} {\bibfnamefont {H.-C.}\ \bibnamefont {Yang}},
  \bibinfo {author} {\bibfnamefont {B.-J.}\ \bibnamefont {Zhang}}, \bibinfo
  {author} {\bibfnamefont {K.}~\bibnamefont {Liu}}, \ and\ \bibinfo {author}
  {\bibfnamefont {Z.-Y.}\ \bibnamefont {Lu}},\ }\href {\doibase
  10.1103/PhysRevB.93.235142} {\bibfield  {journal} {\bibinfo  {journal} {Phys.
  Rev. B}\ }\textbf {\bibinfo {volume} {93}},\ \bibinfo {pages} {235142}
  (\bibinfo {year} {2016})}\BibitemShut {NoStop}%
\bibitem [{\citenamefont {Niu}\ \emph {et~al.}(2016)\citenamefont {Niu},
  \citenamefont {Xu}, \citenamefont {Bai}, \citenamefont {Song}, \citenamefont
  {Shen}, \citenamefont {Xie}, \citenamefont {Sun}, \citenamefont {Huang},
  \citenamefont {Peets},\ and\ \citenamefont {Feng}}]{PhysRevB.94.165163}%
  \BibitemOpen
  \bibfield  {author} {\bibinfo {author} {\bibfnamefont {X.~H.}\ \bibnamefont
  {Niu}}, \bibinfo {author} {\bibfnamefont {D.~F.}\ \bibnamefont {Xu}},
  \bibinfo {author} {\bibfnamefont {Y.~H.}\ \bibnamefont {Bai}}, \bibinfo
  {author} {\bibfnamefont {Q.}~\bibnamefont {Song}}, \bibinfo {author}
  {\bibfnamefont {X.~P.}\ \bibnamefont {Shen}}, \bibinfo {author}
  {\bibfnamefont {B.~P.}\ \bibnamefont {Xie}}, \bibinfo {author} {\bibfnamefont
  {Z.}~\bibnamefont {Sun}}, \bibinfo {author} {\bibfnamefont {Y.~B.}\
  \bibnamefont {Huang}}, \bibinfo {author} {\bibfnamefont {D.~C.}\ \bibnamefont
  {Peets}}, \ and\ \bibinfo {author} {\bibfnamefont {D.~L.}\ \bibnamefont
  {Feng}},\ }\href {\doibase 10.1103/PhysRevB.94.165163} {\bibfield  {journal}
  {\bibinfo  {journal} {Phys. Rev. B}\ }\textbf {\bibinfo {volume} {94}},\
  \bibinfo {pages} {165163} (\bibinfo {year} {2016})}\BibitemShut {NoStop}%
\bibitem [{\citenamefont {Ghimire}\ \emph {et~al.}(2016)\citenamefont
  {Ghimire}, \citenamefont {Botana}, \citenamefont {Phelan}, \citenamefont
  {Zheng},\ and\ \citenamefont {Mitchell}}]{ghimire2016magnetotransport}%
  \BibitemOpen
  \bibfield  {author} {\bibinfo {author} {\bibfnamefont {N.}~\bibnamefont
  {Ghimire}}, \bibinfo {author} {\bibfnamefont {A.}~\bibnamefont {Botana}},
  \bibinfo {author} {\bibfnamefont {D.}~\bibnamefont {Phelan}}, \bibinfo
  {author} {\bibfnamefont {H.}~\bibnamefont {Zheng}}, \ and\ \bibinfo {author}
  {\bibfnamefont {J.}~\bibnamefont {Mitchell}},\ }\href@noop {} {\bibfield
  {journal} {\bibinfo  {journal} {J. Phys.: Condens. Matter}\ }\textbf
  {\bibinfo {volume} {28}},\ \bibinfo {pages} {235601} (\bibinfo {year}
  {2016})}\BibitemShut {NoStop}%
\bibitem [{\citenamefont {Yu}\ \emph {et~al.}(2017)\citenamefont {Yu},
  \citenamefont {Wang}, \citenamefont {Lou}, \citenamefont {Guo}, \citenamefont
  {Xu}, \citenamefont {Liu}, \citenamefont {Wang},\ and\ \citenamefont
  {Xia}}]{Yu2017Magnetoresistance}%
  \BibitemOpen
  \bibfield  {author} {\bibinfo {author} {\bibfnamefont {Q.-H.}\ \bibnamefont
  {Yu}}, \bibinfo {author} {\bibfnamefont {Y.-Y.}\ \bibnamefont {Wang}},
  \bibinfo {author} {\bibfnamefont {R.}~\bibnamefont {Lou}}, \bibinfo {author}
  {\bibfnamefont {P.-J.}\ \bibnamefont {Guo}}, \bibinfo {author} {\bibfnamefont
  {s.}~\bibnamefont {Xu}}, \bibinfo {author} {\bibfnamefont {K.}~\bibnamefont
  {Liu}}, \bibinfo {author} {\bibfnamefont {S.}~\bibnamefont {Wang}}, \ and\
  \bibinfo {author} {\bibfnamefont {T.-L.}\ \bibnamefont {Xia}},\ }\href@noop
  {} {\bibfield  {journal} {\bibinfo  {journal} {EPL}\ }\textbf {\bibinfo
  {volume} {119}},\ \bibinfo {pages} {17002} (\bibinfo {year}
  {2017})}\BibitemShut {NoStop}%
\bibitem [{\citenamefont {Pavlosiuk}\ \emph {et~al.}(2016)\citenamefont
  {Pavlosiuk}, \citenamefont {Swatek},\ and\ \citenamefont
  {Wi{\'s}niewski}}]{pavlosiuk2016giant}%
  \BibitemOpen
  \bibfield  {author} {\bibinfo {author} {\bibfnamefont {O.}~\bibnamefont
  {Pavlosiuk}}, \bibinfo {author} {\bibfnamefont {P.}~\bibnamefont {Swatek}}, \
  and\ \bibinfo {author} {\bibfnamefont {P.}~\bibnamefont {Wi{\'s}niewski}},\
  }\href@noop {} {\bibfield  {journal} {\bibinfo  {journal} {Sci Rep.}\
  }\textbf {\bibinfo {volume} {6}},\ \bibinfo {pages} {38691} (\bibinfo {year}
  {2016})}\BibitemShut {NoStop}%
\bibitem [{\citenamefont {He}\ \emph {et~al.}(2016)\citenamefont {He},
  \citenamefont {Zhang}, \citenamefont {Ghimire}, \citenamefont {Liang},
  \citenamefont {Jia}, \citenamefont {Jiang}, \citenamefont {Tang},
  \citenamefont {Chen}, \citenamefont {He}, \citenamefont {Mo}, \citenamefont
  {Hwang}, \citenamefont {Hashimoto}, \citenamefont {Lu}, \citenamefont
  {Moritz}, \citenamefont {Devereaux}, \citenamefont {Chen}, \citenamefont
  {Mitchell},\ and\ \citenamefont {Shen}}]{PhysRevLett.117.267201}%
  \BibitemOpen
  \bibfield  {author} {\bibinfo {author} {\bibfnamefont {J.}~\bibnamefont
  {He}}, \bibinfo {author} {\bibfnamefont {C.}~\bibnamefont {Zhang}}, \bibinfo
  {author} {\bibfnamefont {N.~J.}\ \bibnamefont {Ghimire}}, \bibinfo {author}
  {\bibfnamefont {T.}~\bibnamefont {Liang}}, \bibinfo {author} {\bibfnamefont
  {C.}~\bibnamefont {Jia}}, \bibinfo {author} {\bibfnamefont {J.}~\bibnamefont
  {Jiang}}, \bibinfo {author} {\bibfnamefont {S.}~\bibnamefont {Tang}},
  \bibinfo {author} {\bibfnamefont {S.}~\bibnamefont {Chen}}, \bibinfo {author}
  {\bibfnamefont {Y.}~\bibnamefont {He}}, \bibinfo {author} {\bibfnamefont
  {S.-K.}\ \bibnamefont {Mo}}, \bibinfo {author} {\bibfnamefont {C.~C.}\
  \bibnamefont {Hwang}}, \bibinfo {author} {\bibfnamefont {M.}~\bibnamefont
  {Hashimoto}}, \bibinfo {author} {\bibfnamefont {D.~H.}\ \bibnamefont {Lu}},
  \bibinfo {author} {\bibfnamefont {B.}~\bibnamefont {Moritz}}, \bibinfo
  {author} {\bibfnamefont {T.~P.}\ \bibnamefont {Devereaux}}, \bibinfo {author}
  {\bibfnamefont {Y.~L.}\ \bibnamefont {Chen}}, \bibinfo {author}
  {\bibfnamefont {J.~F.}\ \bibnamefont {Mitchell}}, \ and\ \bibinfo {author}
  {\bibfnamefont {Z.-X.}\ \bibnamefont {Shen}},\ }\href {\doibase
  10.1103/PhysRevLett.117.267201} {\bibfield  {journal} {\bibinfo  {journal}
  {Phys. Rev. Lett.}\ }\textbf {\bibinfo {volume} {117}},\ \bibinfo {pages}
  {267201} (\bibinfo {year} {2016})}\BibitemShut {NoStop}%
\bibitem [{\citenamefont {Xu}\ \emph {et~al.}(2017)\citenamefont {Xu},
  \citenamefont {Ghimire}, \citenamefont {Jiang}, \citenamefont {Xiao},
  \citenamefont {Botana}, \citenamefont {Wang}, \citenamefont {Hao},
  \citenamefont {Pearson},\ and\ \citenamefont {Kwok}}]{PhysRevB.96.075159}%
  \BibitemOpen
  \bibfield  {author} {\bibinfo {author} {\bibfnamefont {J.}~\bibnamefont
  {Xu}}, \bibinfo {author} {\bibfnamefont {N.~J.}\ \bibnamefont {Ghimire}},
  \bibinfo {author} {\bibfnamefont {J.~S.}\ \bibnamefont {Jiang}}, \bibinfo
  {author} {\bibfnamefont {Z.~L.}\ \bibnamefont {Xiao}}, \bibinfo {author}
  {\bibfnamefont {A.~S.}\ \bibnamefont {Botana}}, \bibinfo {author}
  {\bibfnamefont {Y.~L.}\ \bibnamefont {Wang}}, \bibinfo {author}
  {\bibfnamefont {Y.}~\bibnamefont {Hao}}, \bibinfo {author} {\bibfnamefont
  {J.~E.}\ \bibnamefont {Pearson}}, \ and\ \bibinfo {author} {\bibfnamefont
  {W.~K.}\ \bibnamefont {Kwok}},\ }\href {\doibase 10.1103/PhysRevB.96.075159}
  {\bibfield  {journal} {\bibinfo  {journal} {Phys. Rev. B}\ }\textbf {\bibinfo
  {volume} {96}},\ \bibinfo {pages} {075159} (\bibinfo {year}
  {2017})}\BibitemShut {NoStop}%
\bibitem [{\citenamefont {Alidoust}\ \emph {et~al.}(2016)\citenamefont
  {Alidoust}, \citenamefont {Alexandradinata}, \citenamefont {Xu},
  \citenamefont {Belopolski}, \citenamefont {Kushwaha}, \citenamefont {Zeng},
  \citenamefont {Neupane}, \citenamefont {Bian}, \citenamefont {Liu},
  \citenamefont {Sanchez} \emph {et~al.}}]{alidoust2016new}%
  \BibitemOpen
  \bibfield  {author} {\bibinfo {author} {\bibfnamefont {N.}~\bibnamefont
  {Alidoust}}, \bibinfo {author} {\bibfnamefont {A.}~\bibnamefont
  {Alexandradinata}}, \bibinfo {author} {\bibfnamefont {S.-Y.}\ \bibnamefont
  {Xu}}, \bibinfo {author} {\bibfnamefont {I.}~\bibnamefont {Belopolski}},
  \bibinfo {author} {\bibfnamefont {S.~K.}\ \bibnamefont {Kushwaha}}, \bibinfo
  {author} {\bibfnamefont {M.}~\bibnamefont {Zeng}}, \bibinfo {author}
  {\bibfnamefont {M.}~\bibnamefont {Neupane}}, \bibinfo {author} {\bibfnamefont
  {G.}~\bibnamefont {Bian}}, \bibinfo {author} {\bibfnamefont {C.}~\bibnamefont
  {Liu}}, \bibinfo {author} {\bibfnamefont {D.~S.}\ \bibnamefont {Sanchez}},
  \emph {et~al.},\ }\href@noop {} {\bibfield  {journal} {\bibinfo  {journal}
  {arXiv preprint arXiv:1604.08571}\ } (\bibinfo {year} {2016})}\BibitemShut
  {NoStop}%
\bibitem [{\citenamefont {Ye}\ \emph {et~al.}(2017)\citenamefont {Ye},
  \citenamefont {Suzuki}, \citenamefont {Wicker},\ and\ \citenamefont
  {Checkelsky}}]{ye2017extreme}%
  \BibitemOpen
  \bibfield  {author} {\bibinfo {author} {\bibfnamefont {L.}~\bibnamefont
  {Ye}}, \bibinfo {author} {\bibfnamefont {T.}~\bibnamefont {Suzuki}}, \bibinfo
  {author} {\bibfnamefont {C.~R.}\ \bibnamefont {Wicker}}, \ and\ \bibinfo
  {author} {\bibfnamefont {J.~G.}\ \bibnamefont {Checkelsky}},\ }\href@noop {}
  {\bibfield  {journal} {\bibinfo  {journal} {arXiv preprint arXiv:1704.04226}\
  } (\bibinfo {year} {2017})}\BibitemShut {NoStop}%
\bibitem [{\citenamefont {Oinuma}\ \emph {et~al.}(2017)\citenamefont {Oinuma},
  \citenamefont {Souma}, \citenamefont {Takane}, \citenamefont {Nakamura},
  \citenamefont {Nakayama}, \citenamefont {Mitsuhashi}, \citenamefont {Horiba},
  \citenamefont {Kumigashira}, \citenamefont {Yoshida}, \citenamefont {Ochiai},
  \citenamefont {Takahashi},\ and\ \citenamefont {Sato}}]{PhysRevB.96.041120}%
  \BibitemOpen
  \bibfield  {author} {\bibinfo {author} {\bibfnamefont {H.}~\bibnamefont
  {Oinuma}}, \bibinfo {author} {\bibfnamefont {S.}~\bibnamefont {Souma}},
  \bibinfo {author} {\bibfnamefont {D.}~\bibnamefont {Takane}}, \bibinfo
  {author} {\bibfnamefont {T.}~\bibnamefont {Nakamura}}, \bibinfo {author}
  {\bibfnamefont {K.}~\bibnamefont {Nakayama}}, \bibinfo {author}
  {\bibfnamefont {T.}~\bibnamefont {Mitsuhashi}}, \bibinfo {author}
  {\bibfnamefont {K.}~\bibnamefont {Horiba}}, \bibinfo {author} {\bibfnamefont
  {H.}~\bibnamefont {Kumigashira}}, \bibinfo {author} {\bibfnamefont
  {M.}~\bibnamefont {Yoshida}}, \bibinfo {author} {\bibfnamefont
  {A.}~\bibnamefont {Ochiai}}, \bibinfo {author} {\bibfnamefont
  {T.}~\bibnamefont {Takahashi}}, \ and\ \bibinfo {author} {\bibfnamefont
  {T.}~\bibnamefont {Sato}},\ }\href {\doibase 10.1103/PhysRevB.96.041120}
  {\bibfield  {journal} {\bibinfo  {journal} {Phys. Rev. B}\ }\textbf {\bibinfo
  {volume} {96}},\ \bibinfo {pages} {041120} (\bibinfo {year}
  {2017})}\BibitemShut {NoStop}%
\bibitem [{\citenamefont {Kuroda}\ \emph {et~al.}(2017)\citenamefont {Kuroda},
  \citenamefont {Ochi}, \citenamefont {Suzuki}, \citenamefont {Hirayama},
  \citenamefont {Nakayama}, \citenamefont {Noguchi}, \citenamefont {Bareille},
  \citenamefont {Akebi}, \citenamefont {Kunisada}, \citenamefont {Muro} \emph
  {et~al.}}]{kuroda2017experimental}%
  \BibitemOpen
  \bibfield  {author} {\bibinfo {author} {\bibfnamefont {K.}~\bibnamefont
  {Kuroda}}, \bibinfo {author} {\bibfnamefont {M.}~\bibnamefont {Ochi}},
  \bibinfo {author} {\bibfnamefont {H.}~\bibnamefont {Suzuki}}, \bibinfo
  {author} {\bibfnamefont {M.}~\bibnamefont {Hirayama}}, \bibinfo {author}
  {\bibfnamefont {M.}~\bibnamefont {Nakayama}}, \bibinfo {author}
  {\bibfnamefont {R.}~\bibnamefont {Noguchi}}, \bibinfo {author} {\bibfnamefont
  {C.}~\bibnamefont {Bareille}}, \bibinfo {author} {\bibfnamefont
  {S.}~\bibnamefont {Akebi}}, \bibinfo {author} {\bibfnamefont
  {S.}~\bibnamefont {Kunisada}}, \bibinfo {author} {\bibfnamefont
  {T.}~\bibnamefont {Muro}},  \emph {et~al.},\ }\href@noop {} {\bibfield
  {journal} {\bibinfo  {journal} {arXiv preprint arXiv:1707.06500}\ } (\bibinfo
  {year} {2017})}\BibitemShut {NoStop}%
\bibitem [{\citenamefont {Wakeham}\ \emph {et~al.}(2016)\citenamefont
  {Wakeham}, \citenamefont {Bauer}, \citenamefont {Neupane},\ and\
  \citenamefont {Ronning}}]{PhysRevB.93.205152}%
  \BibitemOpen
  \bibfield  {author} {\bibinfo {author} {\bibfnamefont {N.}~\bibnamefont
  {Wakeham}}, \bibinfo {author} {\bibfnamefont {E.~D.}\ \bibnamefont {Bauer}},
  \bibinfo {author} {\bibfnamefont {M.}~\bibnamefont {Neupane}}, \ and\
  \bibinfo {author} {\bibfnamefont {F.}~\bibnamefont {Ronning}},\ }\href
  {\doibase 10.1103/PhysRevB.93.205152} {\bibfield  {journal} {\bibinfo
  {journal} {Phys. Rev. B}\ }\textbf {\bibinfo {volume} {93}},\ \bibinfo
  {pages} {205152} (\bibinfo {year} {2016})}\BibitemShut {NoStop}%
\bibitem [{\citenamefont {Neupane}\ \emph {et~al.}(2016)\citenamefont
  {Neupane}, \citenamefont {Hosen}, \citenamefont {Belopolski}, \citenamefont
  {Wakeham}, \citenamefont {Dimitri}, \citenamefont {Dhakal}, \citenamefont
  {Zhu}, \citenamefont {Hasan}, \citenamefont {Bauer},\ and\ \citenamefont
  {Ronning}}]{neupane2016observation}%
  \BibitemOpen
  \bibfield  {author} {\bibinfo {author} {\bibfnamefont {M.}~\bibnamefont
  {Neupane}}, \bibinfo {author} {\bibfnamefont {M.~M.}\ \bibnamefont {Hosen}},
  \bibinfo {author} {\bibfnamefont {I.}~\bibnamefont {Belopolski}}, \bibinfo
  {author} {\bibfnamefont {N.}~\bibnamefont {Wakeham}}, \bibinfo {author}
  {\bibfnamefont {K.}~\bibnamefont {Dimitri}}, \bibinfo {author} {\bibfnamefont
  {N.}~\bibnamefont {Dhakal}}, \bibinfo {author} {\bibfnamefont {J.-X.}\
  \bibnamefont {Zhu}}, \bibinfo {author} {\bibfnamefont {M.~Z.}\ \bibnamefont
  {Hasan}}, \bibinfo {author} {\bibfnamefont {E.~D.}\ \bibnamefont {Bauer}}, \
  and\ \bibinfo {author} {\bibfnamefont {F.}~\bibnamefont {Ronning}},\
  }\href@noop {} {\bibfield  {journal} {\bibinfo  {journal} {J. Phys.-Condes.
  Matter}\ }\textbf {\bibinfo {volume} {28}},\ \bibinfo {pages} {23LT02}
  (\bibinfo {year} {2016})}\BibitemShut {NoStop}%
\bibitem [{\citenamefont {Wu}\ \emph {et~al.}(2017)\citenamefont {Wu},
  \citenamefont {Lee}, \citenamefont {Kong}, \citenamefont {Mou}, \citenamefont
  {Jiang}, \citenamefont {Huang}, \citenamefont {Bud'ko}, \citenamefont
  {Canfield},\ and\ \citenamefont {Kaminski}}]{PhysRevB.96.035134}%
  \BibitemOpen
  \bibfield  {author} {\bibinfo {author} {\bibfnamefont {Y.}~\bibnamefont
  {Wu}}, \bibinfo {author} {\bibfnamefont {Y.}~\bibnamefont {Lee}}, \bibinfo
  {author} {\bibfnamefont {T.}~\bibnamefont {Kong}}, \bibinfo {author}
  {\bibfnamefont {D.}~\bibnamefont {Mou}}, \bibinfo {author} {\bibfnamefont
  {R.}~\bibnamefont {Jiang}}, \bibinfo {author} {\bibfnamefont
  {L.}~\bibnamefont {Huang}}, \bibinfo {author} {\bibfnamefont {S.~L.}\
  \bibnamefont {Bud'ko}}, \bibinfo {author} {\bibfnamefont {P.~C.}\
  \bibnamefont {Canfield}}, \ and\ \bibinfo {author} {\bibfnamefont
  {A.}~\bibnamefont {Kaminski}},\ }\href {\doibase 10.1103/PhysRevB.96.035134}
  {\bibfield  {journal} {\bibinfo  {journal} {Phys. Rev. B}\ }\textbf {\bibinfo
  {volume} {96}},\ \bibinfo {pages} {035134} (\bibinfo {year}
  {2017})}\BibitemShut {NoStop}%
\bibitem [{\citenamefont {Ali}\ \emph {et~al.}(2014)\citenamefont {Ali},
  \citenamefont {Xiong}, \citenamefont {Flynn}, \citenamefont {Tao},
  \citenamefont {Gibson}, \citenamefont {Schoop}, \citenamefont {Liang},
  \citenamefont {Haldolaarachchige}, \citenamefont {Hirschberger},
  \citenamefont {Ong},\ and\ \citenamefont {Cava}}]{ali2014large}%
  \BibitemOpen
  \bibfield  {author} {\bibinfo {author} {\bibfnamefont {M.~N.}\ \bibnamefont
  {Ali}}, \bibinfo {author} {\bibfnamefont {J.}~\bibnamefont {Xiong}}, \bibinfo
  {author} {\bibfnamefont {S.}~\bibnamefont {Flynn}}, \bibinfo {author}
  {\bibfnamefont {J.}~\bibnamefont {Tao}}, \bibinfo {author} {\bibfnamefont
  {Q.~D.}\ \bibnamefont {Gibson}}, \bibinfo {author} {\bibfnamefont {L.~M.}\
  \bibnamefont {Schoop}}, \bibinfo {author} {\bibfnamefont {T.}~\bibnamefont
  {Liang}}, \bibinfo {author} {\bibfnamefont {N.}~\bibnamefont
  {Haldolaarachchige}}, \bibinfo {author} {\bibfnamefont {M.}~\bibnamefont
  {Hirschberger}}, \bibinfo {author} {\bibfnamefont {N.~P.}\ \bibnamefont
  {Ong}}, \ and\ \bibinfo {author} {\bibfnamefont {R.~J.}\ \bibnamefont
  {Cava}},\ }\href@noop {} {\bibfield  {journal} {\bibinfo  {journal} {Nature}\
  }\textbf {\bibinfo {volume} {514}},\ \bibinfo {pages} {205} (\bibinfo {year}
  {2014})}\BibitemShut {NoStop}%
\bibitem [{\citenamefont {Wang}\ \emph {et~al.}(2015)\citenamefont {Wang},
  \citenamefont {Thoutam}, \citenamefont {Xiao}, \citenamefont {Hu},
  \citenamefont {Das}, \citenamefont {Mao}, \citenamefont {Wei}, \citenamefont
  {Divan}, \citenamefont {Luican-Mayer}, \citenamefont {Crabtree},\ and\
  \citenamefont {Kwok}}]{PhysRevB.92.180402}%
  \BibitemOpen
  \bibfield  {author} {\bibinfo {author} {\bibfnamefont {Y.~L.}\ \bibnamefont
  {Wang}}, \bibinfo {author} {\bibfnamefont {L.~R.}\ \bibnamefont {Thoutam}},
  \bibinfo {author} {\bibfnamefont {Z.~L.}\ \bibnamefont {Xiao}}, \bibinfo
  {author} {\bibfnamefont {J.}~\bibnamefont {Hu}}, \bibinfo {author}
  {\bibfnamefont {S.}~\bibnamefont {Das}}, \bibinfo {author} {\bibfnamefont
  {Z.~Q.}\ \bibnamefont {Mao}}, \bibinfo {author} {\bibfnamefont
  {J.}~\bibnamefont {Wei}}, \bibinfo {author} {\bibfnamefont {R.}~\bibnamefont
  {Divan}}, \bibinfo {author} {\bibfnamefont {A.}~\bibnamefont {Luican-Mayer}},
  \bibinfo {author} {\bibfnamefont {G.~W.}\ \bibnamefont {Crabtree}}, \ and\
  \bibinfo {author} {\bibfnamefont {W.~K.}\ \bibnamefont {Kwok}},\ }\href
  {\doibase 10.1103/PhysRevB.92.180402} {\bibfield  {journal} {\bibinfo
  {journal} {Phys. Rev. B}\ }\textbf {\bibinfo {volume} {92}},\ \bibinfo
  {pages} {180402} (\bibinfo {year} {2015})}\BibitemShut {NoStop}%
\bibitem [{\citenamefont {Wang}\ \emph {et~al.}(2016)\citenamefont {Wang},
  \citenamefont {Yu}, \citenamefont {Guo}, \citenamefont {Liu},\ and\
  \citenamefont {Xia}}]{PhysRevB.94.041103}%
  \BibitemOpen
  \bibfield  {author} {\bibinfo {author} {\bibfnamefont {Y.-Y.}\ \bibnamefont
  {Wang}}, \bibinfo {author} {\bibfnamefont {Q.-H.}\ \bibnamefont {Yu}},
  \bibinfo {author} {\bibfnamefont {P.-J.}\ \bibnamefont {Guo}}, \bibinfo
  {author} {\bibfnamefont {K.}~\bibnamefont {Liu}}, \ and\ \bibinfo {author}
  {\bibfnamefont {T.-L.}\ \bibnamefont {Xia}},\ }\href {\doibase
  10.1103/PhysRevB.94.041103} {\bibfield  {journal} {\bibinfo  {journal} {Phys.
  Rev. B}\ }\textbf {\bibinfo {volume} {94}},\ \bibinfo {pages} {041103}
  (\bibinfo {year} {2016})}\BibitemShut {NoStop}%
\bibitem [{\citenamefont {Yuan}\ \emph {et~al.}(2016)\citenamefont {Yuan},
  \citenamefont {Lu}, \citenamefont {Liu}, \citenamefont {Wang},\ and\
  \citenamefont {Jia}}]{PhysRevB.93.184405}%
  \BibitemOpen
  \bibfield  {author} {\bibinfo {author} {\bibfnamefont {Z.}~\bibnamefont
  {Yuan}}, \bibinfo {author} {\bibfnamefont {H.}~\bibnamefont {Lu}}, \bibinfo
  {author} {\bibfnamefont {Y.}~\bibnamefont {Liu}}, \bibinfo {author}
  {\bibfnamefont {J.}~\bibnamefont {Wang}}, \ and\ \bibinfo {author}
  {\bibfnamefont {S.}~\bibnamefont {Jia}},\ }\href {\doibase
  10.1103/PhysRevB.93.184405} {\bibfield  {journal} {\bibinfo  {journal} {Phys.
  Rev. B}\ }\textbf {\bibinfo {volume} {93}},\ \bibinfo {pages} {184405}
  (\bibinfo {year} {2016})}\BibitemShut {NoStop}%
\bibitem [{\citenamefont {Wu}\ \emph {et~al.}(2016{\natexlab{b}})\citenamefont
  {Wu}, \citenamefont {Liao}, \citenamefont {Yi}, \citenamefont {Wang},
  \citenamefont {Li}, \citenamefont {Weng}, \citenamefont {Shi}, \citenamefont
  {Li}, \citenamefont {Luo}, \citenamefont {Dai} \emph {et~al.}}]{wu2016giant}%
  \BibitemOpen
  \bibfield  {author} {\bibinfo {author} {\bibfnamefont {D.}~\bibnamefont
  {Wu}}, \bibinfo {author} {\bibfnamefont {J.}~\bibnamefont {Liao}}, \bibinfo
  {author} {\bibfnamefont {W.}~\bibnamefont {Yi}}, \bibinfo {author}
  {\bibfnamefont {X.}~\bibnamefont {Wang}}, \bibinfo {author} {\bibfnamefont
  {P.}~\bibnamefont {Li}}, \bibinfo {author} {\bibfnamefont {H.}~\bibnamefont
  {Weng}}, \bibinfo {author} {\bibfnamefont {Y.}~\bibnamefont {Shi}}, \bibinfo
  {author} {\bibfnamefont {Y.}~\bibnamefont {Li}}, \bibinfo {author}
  {\bibfnamefont {J.}~\bibnamefont {Luo}}, \bibinfo {author} {\bibfnamefont
  {X.}~\bibnamefont {Dai}},  \emph {et~al.},\ }\href@noop {} {\bibfield
  {journal} {\bibinfo  {journal} {Appl. Phys. Lett.}\ }\textbf {\bibinfo
  {volume} {108}},\ \bibinfo {pages} {042105} (\bibinfo {year}
  {2016}{\natexlab{b}})}\BibitemShut {NoStop}%
\bibitem [{\citenamefont {Luo}\ \emph {et~al.}(2016)\citenamefont {Luo},
  \citenamefont {McDonald}, \citenamefont {Rosa}, \citenamefont {Scott},
  \citenamefont {Wakeham}, \citenamefont {Ghimire}, \citenamefont {Bauer},
  \citenamefont {Thompson},\ and\ \citenamefont {Ronning}}]{luo2016anomalous}%
  \BibitemOpen
  \bibfield  {author} {\bibinfo {author} {\bibfnamefont {Y.}~\bibnamefont
  {Luo}}, \bibinfo {author} {\bibfnamefont {R.}~\bibnamefont {McDonald}},
  \bibinfo {author} {\bibfnamefont {P.}~\bibnamefont {Rosa}}, \bibinfo {author}
  {\bibfnamefont {B.}~\bibnamefont {Scott}}, \bibinfo {author} {\bibfnamefont
  {N.}~\bibnamefont {Wakeham}}, \bibinfo {author} {\bibfnamefont
  {N.}~\bibnamefont {Ghimire}}, \bibinfo {author} {\bibfnamefont
  {E.}~\bibnamefont {Bauer}}, \bibinfo {author} {\bibfnamefont
  {J.}~\bibnamefont {Thompson}}, \ and\ \bibinfo {author} {\bibfnamefont
  {F.}~\bibnamefont {Ronning}},\ }\href@noop {} {\bibfield  {journal} {\bibinfo
   {journal} {Sci Rep}\ }\textbf {\bibinfo {volume} {6}},\ \bibinfo {pages}
  {27294} (\bibinfo {year} {2016})}\BibitemShut {NoStop}%
\bibitem [{\citenamefont {Shen}\ \emph {et~al.}(2016)\citenamefont {Shen},
  \citenamefont {Deng}, \citenamefont {Kotliar},\ and\ \citenamefont
  {Ni}}]{PhysRevB.93.195119}%
  \BibitemOpen
  \bibfield  {author} {\bibinfo {author} {\bibfnamefont {B.}~\bibnamefont
  {Shen}}, \bibinfo {author} {\bibfnamefont {X.}~\bibnamefont {Deng}}, \bibinfo
  {author} {\bibfnamefont {G.}~\bibnamefont {Kotliar}}, \ and\ \bibinfo
  {author} {\bibfnamefont {N.}~\bibnamefont {Ni}},\ }\href {\doibase
  10.1103/PhysRevB.93.195119} {\bibfield  {journal} {\bibinfo  {journal} {Phys.
  Rev. B}\ }\textbf {\bibinfo {volume} {93}},\ \bibinfo {pages} {195119}
  (\bibinfo {year} {2016})}\BibitemShut {NoStop}%
\bibitem [{\citenamefont {Liang}\ \emph {et~al.}(2015)\citenamefont {Liang},
  \citenamefont {Gibson}, \citenamefont {Ali}, \citenamefont {Liu},
  \citenamefont {Cava},\ and\ \citenamefont {Ong}}]{liang2015ultrahigh}%
  \BibitemOpen
  \bibfield  {author} {\bibinfo {author} {\bibfnamefont {T.}~\bibnamefont
  {Liang}}, \bibinfo {author} {\bibfnamefont {Q.}~\bibnamefont {Gibson}},
  \bibinfo {author} {\bibfnamefont {M.~N.}\ \bibnamefont {Ali}}, \bibinfo
  {author} {\bibfnamefont {M.}~\bibnamefont {Liu}}, \bibinfo {author}
  {\bibfnamefont {R.}~\bibnamefont {Cava}}, \ and\ \bibinfo {author}
  {\bibfnamefont {N.}~\bibnamefont {Ong}},\ }\href@noop {} {\bibfield
  {journal} {\bibinfo  {journal} {Nat. Mater.}\ }\textbf {\bibinfo {volume}
  {14}},\ \bibinfo {pages} {280} (\bibinfo {year} {2015})}\BibitemShut
  {NoStop}%
\bibitem [{\citenamefont {Bl\"ochl}(1994)}]{PhysRevB.50.17953}%
  \BibitemOpen
  \bibfield  {author} {\bibinfo {author} {\bibfnamefont {P.~E.}\ \bibnamefont
  {Bl\"ochl}},\ }\href@noop {} {\bibfield  {journal} {\bibinfo  {journal}
  {Phys. Rev. B}\ }\textbf {\bibinfo {volume} {50}},\ \bibinfo {pages} {17953}
  (\bibinfo {year} {1994})}\BibitemShut {NoStop}%
\bibitem [{\citenamefont {Kresse}\ and\ \citenamefont
  {Joubert}(1999)}]{PhysRevB.59.1758}%
  \BibitemOpen
  \bibfield  {author} {\bibinfo {author} {\bibfnamefont {G.}~\bibnamefont
  {Kresse}}\ and\ \bibinfo {author} {\bibfnamefont {D.}~\bibnamefont
  {Joubert}},\ }\href@noop {} {\bibfield  {journal} {\bibinfo  {journal} {Phys.
  Rev. B}\ }\textbf {\bibinfo {volume} {59}},\ \bibinfo {pages} {1758}
  (\bibinfo {year} {1999})}\BibitemShut {NoStop}%
\bibitem [{\citenamefont {Kresse}\ and\ \citenamefont
  {Hafner}(1993)}]{PhysRevB.47.558}%
  \BibitemOpen
  \bibfield  {author} {\bibinfo {author} {\bibfnamefont {G.}~\bibnamefont
  {Kresse}}\ and\ \bibinfo {author} {\bibfnamefont {J.}~\bibnamefont
  {Hafner}},\ }\href@noop {} {\bibfield  {journal} {\bibinfo  {journal} {Phys.
  Rev. B}\ }\textbf {\bibinfo {volume} {47}},\ \bibinfo {pages} {558} (\bibinfo
  {year} {1993})}\BibitemShut {NoStop}%
\bibitem [{\citenamefont {Kresse}\ and\ \citenamefont
  {Furthm{\"u}ller}(1996)}]{kresse1996efficiency}%
  \BibitemOpen
  \bibfield  {author} {\bibinfo {author} {\bibfnamefont {G.}~\bibnamefont
  {Kresse}}\ and\ \bibinfo {author} {\bibfnamefont {J.}~\bibnamefont
  {Furthm{\"u}ller}},\ }\href@noop {} {\bibfield  {journal} {\bibinfo
  {journal} {Comp. Mater. Sci.}\ }\textbf {\bibinfo {volume} {6}},\ \bibinfo
  {pages} {15} (\bibinfo {year} {1996})}\BibitemShut {NoStop}%
\bibitem [{\citenamefont {Kresse}\ and\ \citenamefont
  {Furthm\"uller}(1996)}]{PhysRevB.54.11169}%
  \BibitemOpen
  \bibfield  {author} {\bibinfo {author} {\bibfnamefont {G.}~\bibnamefont
  {Kresse}}\ and\ \bibinfo {author} {\bibfnamefont {J.}~\bibnamefont
  {Furthm\"uller}},\ }\href@noop {} {\bibfield  {journal} {\bibinfo  {journal}
  {Phys. Rev. B}\ }\textbf {\bibinfo {volume} {54}},\ \bibinfo {pages} {11169}
  (\bibinfo {year} {1996})}\BibitemShut {NoStop}%
\bibitem [{\citenamefont {Perdew}\ \emph {et~al.}(1996)\citenamefont {Perdew},
  \citenamefont {Burke},\ and\ \citenamefont
  {Ernzerhof}}]{PhysRevLett.77.3865}%
  \BibitemOpen
  \bibfield  {author} {\bibinfo {author} {\bibfnamefont {J.~P.}\ \bibnamefont
  {Perdew}}, \bibinfo {author} {\bibfnamefont {K.}~\bibnamefont {Burke}}, \
  and\ \bibinfo {author} {\bibfnamefont {M.}~\bibnamefont {Ernzerhof}},\
  }\href@noop {} {\bibfield  {journal} {\bibinfo  {journal} {Phys. Rev. Lett.}\
  }\textbf {\bibinfo {volume} {77}},\ \bibinfo {pages} {3865} (\bibinfo {year}
  {1996})}\BibitemShut {NoStop}%
\bibitem [{\citenamefont {Abdusalyamova}\ \emph {et~al.}(1994)\citenamefont
  {Abdusalyamova}, \citenamefont {Chuiko}, \citenamefont {Golubkov},
  \citenamefont {Popov}, \citenamefont {Parfenova}, \citenamefont {Procofev},\
  and\ \citenamefont {Smirnov}}]{abdusalyamova1994synthesis}%
  \BibitemOpen
  \bibfield  {author} {\bibinfo {author} {\bibfnamefont {M.}~\bibnamefont
  {Abdusalyamova}}, \bibinfo {author} {\bibfnamefont {A.}~\bibnamefont
  {Chuiko}}, \bibinfo {author} {\bibfnamefont {A.~Y.}\ \bibnamefont
  {Golubkov}}, \bibinfo {author} {\bibfnamefont {S.}~\bibnamefont {Popov}},
  \bibinfo {author} {\bibfnamefont {L.}~\bibnamefont {Parfenova}}, \bibinfo
  {author} {\bibfnamefont {A.}~\bibnamefont {Procofev}}, \ and\ \bibinfo
  {author} {\bibfnamefont {I.}~\bibnamefont {Smirnov}},\ }\href@noop {}
  {\bibfield  {journal} {\bibinfo  {journal} {J. Alloy. Compd.}\ }\textbf
  {\bibinfo {volume} {205}},\ \bibinfo {pages} {107} (\bibinfo {year}
  {1994})}\BibitemShut {NoStop}%
\bibitem [{\citenamefont {Becke}\ and\ \citenamefont
  {Johnson}(2006)}]{becke2006simple}%
  \BibitemOpen
  \bibfield  {author} {\bibinfo {author} {\bibfnamefont {A.~D.}\ \bibnamefont
  {Becke}}\ and\ \bibinfo {author} {\bibfnamefont {E.~R.}\ \bibnamefont
  {Johnson}},\ }\href@noop {} {\bibfield  {journal} {\bibinfo  {journal} {J.
  Chem. Phys.}\ }\textbf {\bibinfo {volume} {124}},\ \bibinfo {pages} {221101}
  (\bibinfo {year} {2006})}\BibitemShut {NoStop}%
\bibitem [{\citenamefont {Tran}\ and\ \citenamefont
  {Blaha}(2009)}]{PhysRevLett.102.226401}%
  \BibitemOpen
  \bibfield  {author} {\bibinfo {author} {\bibfnamefont {F.}~\bibnamefont
  {Tran}}\ and\ \bibinfo {author} {\bibfnamefont {P.}~\bibnamefont {Blaha}},\
  }\href@noop {} {\bibfield  {journal} {\bibinfo  {journal} {Phys. Rev. Lett.}\
  }\textbf {\bibinfo {volume} {102}},\ \bibinfo {pages} {226401} (\bibinfo
  {year} {2009})}\BibitemShut {NoStop}%
\bibitem [{\citenamefont {Marzari}\ and\ \citenamefont
  {Vanderbilt}(1997)}]{PhysRevB.56.12847}%
  \BibitemOpen
  \bibfield  {author} {\bibinfo {author} {\bibfnamefont {N.}~\bibnamefont
  {Marzari}}\ and\ \bibinfo {author} {\bibfnamefont {D.}~\bibnamefont
  {Vanderbilt}},\ }\href@noop {} {\bibfield  {journal} {\bibinfo  {journal}
  {Phys. Rev. B}\ }\textbf {\bibinfo {volume} {56}},\ \bibinfo {pages} {12847}
  (\bibinfo {year} {1997})}\BibitemShut {NoStop}%
\bibitem [{\citenamefont {Souza}\ \emph {et~al.}(2001)\citenamefont {Souza},
  \citenamefont {Marzari},\ and\ \citenamefont
  {Vanderbilt}}]{PhysRevB.65.035109}%
  \BibitemOpen
  \bibfield  {author} {\bibinfo {author} {\bibfnamefont {I.}~\bibnamefont
  {Souza}}, \bibinfo {author} {\bibfnamefont {N.}~\bibnamefont {Marzari}}, \
  and\ \bibinfo {author} {\bibfnamefont {D.}~\bibnamefont {Vanderbilt}},\
  }\href@noop {} {\bibfield  {journal} {\bibinfo  {journal} {Phys. Rev. B}\
  }\textbf {\bibinfo {volume} {65}},\ \bibinfo {pages} {035109} (\bibinfo
  {year} {2001})}\BibitemShut {NoStop}%
\bibitem [{\citenamefont {Nimori}\ \emph {et~al.}(1995)\citenamefont {Nimori},
  \citenamefont {Kido}, \citenamefont {Li},\ and\ \citenamefont
  {Suzuki}}]{nimori1995haas}%
  \BibitemOpen
  \bibfield  {author} {\bibinfo {author} {\bibfnamefont {S.}~\bibnamefont
  {Nimori}}, \bibinfo {author} {\bibfnamefont {G.}~\bibnamefont {Kido}},
  \bibinfo {author} {\bibfnamefont {D.}~\bibnamefont {Li}}, \ and\ \bibinfo
  {author} {\bibfnamefont {T.}~\bibnamefont {Suzuki}},\ }\href@noop {}
  {\bibfield  {journal} {\bibinfo  {journal} {Physica B}\ }\textbf {\bibinfo
  {volume} {211}},\ \bibinfo {pages} {148} (\bibinfo {year}
  {1995})}\BibitemShut {NoStop}%
\bibitem [{\citenamefont {Cooper}\ and\ \citenamefont
  {Vogt}(1970)}]{PhysRevB.1.1211}%
  \BibitemOpen
  \bibfield  {author} {\bibinfo {author} {\bibfnamefont {B.~R.}\ \bibnamefont
  {Cooper}}\ and\ \bibinfo {author} {\bibfnamefont {O.}~\bibnamefont {Vogt}},\
  }\href {\doibase 10.1103/PhysRevB.1.1211} {\bibfield  {journal} {\bibinfo
  {journal} {Phys. Rev. B}\ }\textbf {\bibinfo {volume} {1}},\ \bibinfo {pages}
  {1211} (\bibinfo {year} {1970})}\BibitemShut {NoStop}%
\bibitem [{\citenamefont {{\=O}nuki}\ \emph {et~al.}(2014)\citenamefont
  {{\=O}nuki}, \citenamefont {Nakamura}, \citenamefont {Uejo}, \citenamefont
  {Teruya}, \citenamefont {Hedo}, \citenamefont {Nakama}, \citenamefont
  {Honda},\ and\ \citenamefont {Harima}}]{onuki2014chiral}%
  \BibitemOpen
  \bibfield  {author} {\bibinfo {author} {\bibfnamefont {Y.}~\bibnamefont
  {{\=O}nuki}}, \bibinfo {author} {\bibfnamefont {A.}~\bibnamefont {Nakamura}},
  \bibinfo {author} {\bibfnamefont {T.}~\bibnamefont {Uejo}}, \bibinfo {author}
  {\bibfnamefont {A.}~\bibnamefont {Teruya}}, \bibinfo {author} {\bibfnamefont
  {M.}~\bibnamefont {Hedo}}, \bibinfo {author} {\bibfnamefont {T.}~\bibnamefont
  {Nakama}}, \bibinfo {author} {\bibfnamefont {F.}~\bibnamefont {Honda}}, \
  and\ \bibinfo {author} {\bibfnamefont {H.}~\bibnamefont {Harima}},\
  }\href@noop {} {\bibfield  {journal} {\bibinfo  {journal} {J. Phys. Soc.
  Jpn.}\ }\textbf {\bibinfo {volume} {83}},\ \bibinfo {pages} {061018}
  (\bibinfo {year} {2014})}\BibitemShut {NoStop}%
\bibitem [{\citenamefont {Zhang}\ \emph {et~al.}(2011)\citenamefont {Zhang},
  \citenamefont {Richard}, \citenamefont {Qian}, \citenamefont {Xu},
  \citenamefont {Dai},\ and\ \citenamefont {Ding}}]{zhang2011precise}%
  \BibitemOpen
  \bibfield  {author} {\bibinfo {author} {\bibfnamefont {P.}~\bibnamefont
  {Zhang}}, \bibinfo {author} {\bibfnamefont {P.}~\bibnamefont {Richard}},
  \bibinfo {author} {\bibfnamefont {T.}~\bibnamefont {Qian}}, \bibinfo {author}
  {\bibfnamefont {Y.-M.}\ \bibnamefont {Xu}}, \bibinfo {author} {\bibfnamefont
  {X.}~\bibnamefont {Dai}}, \ and\ \bibinfo {author} {\bibfnamefont
  {H.}~\bibnamefont {Ding}},\ }\href@noop {} {\bibfield  {journal} {\bibinfo
  {journal} {Rev. Sci. Instrum.}\ }\textbf {\bibinfo {volume} {82}},\ \bibinfo
  {pages} {043712} (\bibinfo {year} {2011})}\BibitemShut {NoStop}%
\bibitem [{\citenamefont {Kokalj}(2003)}]{kokalj2003computer}%
  \BibitemOpen
  \bibfield  {author} {\bibinfo {author} {\bibfnamefont {A.}~\bibnamefont
  {Kokalj}},\ }\href@noop {} {\bibfield  {journal} {\bibinfo  {journal} {Comp.
  Mater. Sci.}\ }\textbf {\bibinfo {volume} {28}},\ \bibinfo {pages} {155}
  (\bibinfo {year} {2003})}\BibitemShut {NoStop}%
\end{thebibliography}%
\end{document}